\documentclass[twocolumn,nobibnotes,prx,unsortedaddress]{revtex4-1}
\usepackage[english]{babel}
\usepackage[T1]{fontenc} 
\usepackage[utf8]{inputenc}

\usepackage{graphicx}
\usepackage{dcolumn}
\usepackage{bm}
\usepackage{placeins}
\usepackage{color}
\usepackage[usenames,dvipsnames,svgnames,table]{xcolor}
\usepackage{float}
\usepackage{amsmath}
\usepackage{amssymb}
\usepackage{mathtools} 
\usepackage[colorlinks=true,linkcolor=blue,urlcolor=blue,citecolor=blue]{hyperref}
\usepackage{tikz}
\usetikzlibrary{shapes,patterns}
\usetikzlibrary{shapes.geometric}

\newcommand{\etal}{\text{\emph{et al.} }}
\newcommand{\minutes}{\,\mathrm{min}}
\newcommand{\rads}{\,\mathrm{rad/s}}
\newcommand{\TB}{\textit{TB}}
\newcommand{\degC}{^\circ \mathrm{C}}
\newcommand{\Hz}{\mathrm{Hz}}

\newcommand{\s}{\,\mathrm{s}}
\newcommand{\h}{\,\mathrm{h}}
\newcommand{\Pas}{\,\mathrm{Pa\,s}}
\newcommand{\Pa}{\,\mathrm{Pa}}

\newcommand{\mm}{\,\mathrm{mm}}

\newcommand{\di}{\mathrm{d}}

\newcommand{\Solidline}{\raisebox{2.5pt}{\tikz \draw[very thick] (0pt,0pt) -- (10pt,0pt);}}
\newcommand{\dashedline}{\raisebox{2.5pt}{\tikz \draw[very thick,dashed] (0pt,0pt) -- (10pt,0pt);}}
\newcommand{\solidcircleA}{\raisebox{0.5pt}{\tikz \filldraw[very thick,color=red!,fill=red!, draw=red!] (0pt,0pt) -- (15pt,0pt) (7.5pt,0pt) circle (2.5pt);}}
\newcommand{\dashedcircleA}{\raisebox{0.5pt}{\tikz \filldraw[very thick,color=red!,solid] (0pt,0pt) -- (3pt,0pt) [fill=white!, draw=red]  (7.5pt,0pt) circle (2.5pt)  (12pt,0pt) -- (15pt,0pt);}}

\newcommand{\SolidlineB}{\raisebox{2.5pt}{\tikz \draw[very thick,color=black!50] (0pt,0pt) -- (10pt,0pt);}}
\newcommand{\dashedlineB}{\raisebox{2.5pt}{\tikz \draw[very thick,dashed,color=black!80] (0pt,0pt) -- (10pt,0pt);}}
\definecolor{LightBlue}{rgb}{0, 0.4, 0.8}
\newcommand{\SolidlineC}{\raisebox{2.5pt}{\tikz \draw[very thick,color=LightBlue!] (0pt,0pt) -- (10pt,0pt);}}
\definecolor{DarkBlue}{rgb}{0, 0, 0.6}
\newcommand{\dashedlineC}{\raisebox{2.5pt}{\tikz \draw[very thick,dashed,color=DarkBlue!] (0pt,0pt) -- (10pt,0pt);}}
\newcommand{\solidsquareDw}{\raisebox{0.5pt}{\tikz \filldraw[very thick,color=black!50,solid] (0pt,0pt) -- (5pt,0pt) [fill=white!, draw=black!50]  (5pt,-2.5pt) rectangle (10pt,2.5pt)  (10pt,0pt) -- (15pt,0pt);}}
\newcommand{\solidsquareD}{\raisebox{0.5pt}{\tikz \filldraw[very thick,color=black!50,solid] (0pt,0pt) -- (5pt,0pt) [fill=black!50, draw=black!50]  (5pt,-2.5pt) rectangle (10pt,2.5pt)  (10pt,0pt) -- (15pt,0pt);}}
\newcommand{\dashedsquareD}{\raisebox{0.5pt}{\tikz \filldraw[very thick,color=black!50,solid] (0pt,0pt) -- (3pt,0pt) [fill=white!, draw=black!50]  (5pt,-2.5pt) rectangle (10pt,2.5pt)  (12pt,0pt) -- (15pt,0pt);}}
\newcommand{\dashedcircleD}{\raisebox{0.5pt}{\tikz \filldraw[very thick,color=LightBlue!,solid] (0pt,0pt) -- (3pt,0pt) [fill=white!, draw=LightBlue!]  (7.5pt,0pt) circle (2.5pt)  (12pt,0pt) -- (15pt,0pt);}}
\newcommand{\solidcircleD}{\raisebox{0.5pt}{\tikz \filldraw[very thick,color=LightBlue!,solid] (0pt,0pt) -- (5pt,0pt) [fill=LightBlue!, draw=LightBlue!]  (7.5pt,0pt) circle (2.5pt)  (10.5pt,0pt) -- (15pt,0pt);}}
\definecolor{LightRed}{rgb}{1, 0, 0.2}
\newcommand{\dashedtriangleE}{\raisebox{0.5pt}{\tikz \filldraw[very thick,color=LightRed!,solid] (0pt,0pt) -- (3pt,0pt) [fill=white!, draw=LightRed!]  (5pt,-2.5pt) -- (10pt,-2.5pt) -- (7.5pt,2.5pt) -- cycle (12pt,0pt) -- (15pt,0pt);}}
\newcommand{\solidtriangleE}{\raisebox{0.5pt}{\tikz \filldraw[very thick,color=LightRed!,solid] (0pt,0pt) -- (15pt,0pt) [fill=LightRed!, draw=LightRed!]  (5pt,-2.5pt) -- (10pt,-2.5pt) -- (7.5pt,2.5pt) -- cycle;}}

\newcommand{\solidsquareFw}{\raisebox{0.5pt}{\tikz \filldraw[very thick,color=LightBlue!,solid] (0pt,0pt) -- (5pt,0pt) [fill=white!, draw=LightBlue!]  (5pt,-2.5pt) rectangle (10.5pt,2.5pt)  (10.5pt,0pt) -- (15pt,0pt);}}
\newcommand{\solidsquareF}{\raisebox{0.5pt}{\tikz \filldraw[very thick,color=LightBlue!,solid] (0pt,0pt) -- (5pt,0pt) [fill=LightBlue!, draw=LightBlue!]  (5pt,-2.5pt) rectangle (10pt,2.5pt)  (10pt,0pt) -- (15pt,0pt);}}
\definecolor{Green}{rgb}{0, 0.4, 0}
\newcommand{\soliddiamondEw}{\raisebox{0.5pt}{\tikz \filldraw[very thick,color=Green!,solid] (0pt,0pt) -- (5pt,0pt) [fill=white!, draw=Green!]  (5pt,0pt) -- (7.5pt,-3.5pt) -- (10pt,0pt) -- (7.5pt,3.5pt) -- cycle (10pt,0pt) -- (15pt,0pt);}}
\newcommand{\soliddiamondE}{\raisebox{0.5pt}{\tikz \filldraw[very thick,color=Green!,solid] (0pt,0pt) -- (15pt,0pt) [fill=Green!, draw=Green!]  (5pt,0pt) -- (7.5pt,-3.5pt) -- (10pt,0pt) -- (7.5pt,3.5pt) -- cycle;}}
\newcommand{\solidcircleF}{\raisebox{0.5pt}{\tikz \filldraw[very thick,color=black!80,fill=black!80, draw=black!80] (0pt,0pt) -- (15pt,0pt) (7.5pt,0pt) circle (2.5pt);}}
\newcommand{\solidcircleFw}{\raisebox{0.5pt}{\tikz \filldraw[very thick,color=black!80,fill=white!, draw=black!80] (0pt,0pt) -- (5pt,0pt) (7.5pt,0pt) circle (2.5pt) (10pt,0pt) -- (15pt,0pt);}}

\definecolor{Brick}{rgb}{0.6, 0.2, 0.2}
\newcommand{\SolidlineG}{\raisebox{2.5pt}{\tikz \draw[very thick,color=Brick!] (0pt,0pt) -- (10pt,0pt);}}
\definecolor{Pink}{rgb}{1, 0, 1}

\newcommand{\triangleG}{\raisebox{0.0pt}{\tikz \filldraw[very thick, fill=Pink!, draw=Pink!]  (0pt,-2.5pt) -- (4pt,0pt) -- (0pt,2.5pt) -- cycle;}}
\definecolor{Cyan}{rgb}{0.2, 1, 1}
\newcommand{\circleG}{\raisebox{-1.0pt}{\tikz \filldraw[very thick, fill=Cyan!, draw=Cyan!]  (0pt,0pt) circle (2.5pt);}}
\definecolor{Blue}{rgb}{0, 0, 1}
\newcommand{\squareG}{\raisebox{-1.0pt}{\tikz \filldraw[very thick, fill=Blue!, draw=Blue!]  (0pt,-2.5pt) rectangle (5pt,2.5pt);}}
\newcommand{\diamondG}{\raisebox{-1.0pt}{\tikz \filldraw[very thick, fill=red!, draw=red!]  (5pt,0pt) -- (7.5pt,-2.5pt) -- (10pt,0pt) -- (7.5pt,2.5pt) -- cycle;}}

\newcommand{\triangleHw}{\raisebox{0.0pt}{\tikz \filldraw[very thick, fill=white!, draw=Blue!]  (5pt,-2.5pt) -- (10pt,-2.5pt) -- (7.5pt,2.5pt) -- cycle;}}
\newcommand{\triangleH}{\raisebox{0.0pt}{\tikz \filldraw[very thick, fill=Blue!, draw=Blue!]  (5pt,-2.5pt) -- (10pt,-2.5pt) -- (7.5pt,2.5pt) -- cycle;}}
\definecolor{LightGreen}{rgb}{0.4, 0.6, 0}
\newcommand{\squareH}{\raisebox{0.5pt}{\tikz \filldraw[very thick, fill=LightGreen!, draw=LightGreen!]  (0pt,-2.5pt) rectangle (5pt,2.5pt);}}
\newcommand{\squareHw}{\raisebox{0.5pt}{\tikz \filldraw[very thick, fill=white!, draw=LightGreen!]  (0pt,-2.5pt) rectangle (5pt,2.5pt);}}
\newcommand{\circleH}{\raisebox{-1pt}{\tikz \filldraw[very thick, fill=LightRed!, draw=LightRed!]  (0pt,0pt) circle (2.5pt);}}
\newcommand{\circleHw}{\raisebox{-1pt}{\tikz \filldraw[very thick, fill=white!, draw=LightRed!]  (0pt,0pt) circle (2.5pt);}}
\newcommand{\circleI}{\raisebox{-1pt}{\tikz \filldraw[very thick, fill=black!, draw=black!]  (7.5pt,0pt) circle (2.5pt) (0pt,0pt) -- (1pt,0pt)  (2pt,0pt) -- (3pt,0pt) (4pt,0pt) -- (5pt,0pt) (10pt,0pt) -- (11pt,0pt) (12pt,0pt) -- (13pt,0pt) (14pt,0pt) -- (15pt,0pt);}}
\newcommand{\StarI}{\raisebox{0.0pt}{\tikz{\draw[very thick] (0pt,0pt) -- (3pt,0pt); \node[xshift=5.5pt, yshift=-4.5pt, shape=star, fill=black!, draw, black, star point ratio=3, draw,inner sep=0.8pt,anchor=outer point 3]{}; \draw[very thick,xshift=0.0pt, yshift=0pt](12.5pt,0pt) -- (15.5pt,0pt);}}}

\newcommand{\circleg}{\raisebox{-1.0pt}{\tikz \filldraw[very thick, fill=Green!, draw=Green!] (7.5pt,0pt) circle (2.5pt) (0pt,0pt) -- (1pt,0pt)  (2pt,0pt) -- (3pt,0pt) (4pt,0pt) -- (5pt,0pt) (10pt,0pt) -- (11pt,0pt) (12pt,0pt) -- (13pt,0pt) (14pt,0pt) -- (15pt,0pt);}}
\newcommand{\circleb}{\raisebox{-1.0pt}{\tikz \filldraw[very thick, fill=blue!, draw=blue!]  (7.5pt,0pt) circle (2.5pt) (0pt,0pt) -- (1pt,0pt)  (2pt,0pt) -- (3pt,0pt) (4pt,0pt) -- (5pt,0pt) (10pt,0pt) -- (11pt,0pt) (12pt,0pt) -- (13pt,0pt) (14pt,0pt) -- (15pt,0pt);}}
\newcommand{\circlem}{\raisebox{-1.0pt}{\tikz \filldraw[very thick, fill=magenta!, draw=magenta!]  (7.5pt,0pt) circle (2.5pt) (0pt,0pt) -- (1pt,0pt)  (2pt,0pt) -- (3pt,0pt) (4pt,0pt) -- (5pt,0pt) (10pt,0pt) -- (11pt,0pt) (12pt,0pt) -- (13pt,0pt) (14pt,0pt) -- (15pt,0pt);}}
\newcommand{\circler}{\raisebox{-1.0pt}{\tikz \filldraw[very thick, fill=Red!, draw=Red!]  (7.5pt,0pt) circle (2.5pt) (0pt,0pt) -- (1pt,0pt)  (2pt,0pt) -- (3pt,0pt) (4pt,0pt) -- (5pt,0pt) (10pt,0pt) -- (11pt,0pt) (12pt,0pt) -- (13pt,0pt) (14pt,0pt) -- (15pt,0pt);}}

\newcommand{\solidlineg}{\raisebox{2.5pt}{\tikz \draw[very thick,draw=Green!] (0pt,0pt) -- (10pt,0pt);}}
\newcommand{\solidlineb}{\raisebox{2.5pt}{\tikz \draw[very thick,draw=blue!] (0pt,0pt) -- (10pt,0pt);}}
\newcommand{\solidlinem}{\raisebox{2.5pt}{\tikz \draw[very thick,draw=magenta!] (0pt,0pt) -- (10pt,0pt);}}
\newcommand{\solidliner}{\raisebox{2.5pt}{\tikz \draw[very thick,draw=Red!] (0pt,0pt) -- (10pt,0pt);}}

\begin{document}

\title{Time-Resolved Mechanical Spectroscopy of Soft Materials \\ via Optimally Windowed Chirps}%

\author{Michela Geri\textsuperscript{1}}%
\email[]{mgeri@mit.edu} 
%
\author{Bavand Keshavarz\textsuperscript{1}}%
\email[]{bavand@mit.edu}
\thanks{\\ \textsuperscript{1}These Authors contributed equally}
\affiliation{Department of Mechanical Engineering, Massachusetts Institute of Technology, Cambridge MA 02139, United States}
\author{Thibaut Divoux}%
 \affiliation{Centre de Recherche Paul Pascal, CNRS UMR 5031 - 115 avenue Schweitzer, 33600 Pessac, France}
\affiliation{MultiScale Material Science for Energy and Environment, UMI 3466, CNRS-MIT, 77 Massachusetts Avenue, Cambridge, Massachusetts 02139, USA}
\author{Christian Clasen}%
\affiliation{KU Leuven, Department of Chemical Engineering, 3001 Leuven, Belgium}
\author{Dan J. Curtis}%
\affiliation{Complex Fluids Research Group, College of Engineering, Swansea University, Wales}
\author{Gareth H. McKinley}%
\email[]{gareth@mit.edu}
\affiliation{Department of Mechanical Engineering, Massachusetts Institute of Technology, Cambridge MA 02139, United States}
\date{\today}%


\begin{abstract}
	
	The ability to measure the bulk dynamic behavior of soft materials with combined time- and frequency-resolution is instrumental for improving our fundamental understanding of connections between the microstructural dynamics and the macroscopic mechanical response. Current state-of-the-art techniques are often limited by a compromise between resolution in the time and frequency domain, mainly due to the use of elementary input signals that have not been designed for fast time-evolving systems such as materials undergoing gelation, curing or self-healing.  In this work, we develop an optimized and robust excitation signal for time-resolved mechanical spectroscopy through the introduction of joint frequency- and amplitude-modulated exponential chirps. Inspired by the biosonar signals of bats and dolphins, we optimize the signal profile to maximize the signal-to-noise ratio while minimizing spectral leakage with a carefully-designed modulation of the envelope of the chirp, obtained using a cosine tapered window function. A combined experimental and numerical investigation reveals that there exists an optimal range of window profiles (around $10 \%$ of the total signal length) that minimizes the error with respect to standard single frequency sweep techniques.  The minimum error is set by the noise floor of the instrument, suggesting that the accuracy of an optimally windowed chirp (OWCh) sequence is directly comparable to that achievable with a standard frequency sweep, while the acquisition time can be reduced by up to two orders of magnitude, for comparable spectral content. Finally, we demonstrate the ability of this optimized signal to provide time- and frequency-resolved rheometric data by studying the fast gelation process of an acid-induced protein gel using repeated OWCh pulse sequences. The use of optimally windowed chirps enables a robust time-resolved rheological characterization of a wide range of soft materials undergoing rapid mutation and has the potential to become an invaluable tool for researchers across different disciplines.
	
\end{abstract}

\maketitle

\section{Introduction}

Many soft materials that are of interest for industrial \cite{Mezzenga2005,Chen2010,Gibaud2012} or biomedical \cite{Chan2016,Gong2010}
applications often undergo microstructural changes during their synthesis or assembly as a result of chemical, thermal or mechanical processes. Examples include gelation of polymer \cite{Flory1953,Kavanagh1998,Osada2004}, protein \cite{Bremer1990,Totosaus2002,Leocmach2014,Keshavarz2017} and colloidal gels \cite{Zaccarelli2007,Trappe2000,Lu2013} as well as jamming of glasses \cite{Liu1998,Jaeger2015,Kapnistos2000}. These transient processes are responsible for establishing the final material properties of the system, and the ability to follow their time evolution is essential in the quest to relate changes in the underlying material microstructure to the corresponding rheological response at the macroscopic scale \cite{Kavanagh1998,Chen2010,Nagel2017}. Establishing and quantifying such connections is also of crucial importance in the development and design of the next generation of soft materials, such as disordered colloidal aggregates \cite{Nagel2017}, bio-inspired hydrogels \cite{Grindy2015}, mechanical metamaterials \cite{Bertoldi2017} and jammed granular solids \cite{Brown2010}.

Of particular interest in this regard is the development and evolution of the linear viscoelastic response, i.e., the mechanical relaxation spectrum of a material prior to any damage or plastic deformation is imposed to the initial state of the microstructural components. This response can be fully characterized by the knowledge of the relaxation spectrum $G(t)$ or analogously the complex modulus $G^\star (\omega)$, whose real and imaginary parts correspond, respectively, to
the elastic storage modulus $G'(\omega)$ and the viscous loss modulus $G''(\omega)$ \cite{Bird1987}.

Measuring the material properties for time-evolving systems undergoing structural changes is intrinsically challenging due to the time constraint imposed by the characteristic mutation time of the sample under investigation \cite{Mours1994}. While several techniques are currently available to obtain  detailed quantitative information about the microscopic dynamics of mutating systems, e.g., scattering-based techniques \cite{Adam1988,Kroon1996,Romer2000,Duri2005,Eberle2011}, confocal microscopy \cite{Weeks2000}, nuclear magnetic resonance \cite{Assink1991,Bonn2008}, image correlation microscopy \cite{Larsen2008, Edera2017} and AFM-based spectroscopy \cite{Nia2013}, macroscopic dynamic mechanical properties are particularly challenging to measure with enough time- and frequency-resolution \cite{Kavanagh1998}. This is partially due to the scales at which mechanical forces and deformations are applied,
which usually confines the range of frequencies accessible to a maximum of $100 \Hz$ (with state-of-the-art instruments) and requires experimental time scales on the order of hundreds of seconds to obtain sufficient frequency resolution \cite{Kavanagh1998,Chen2010}.
However this limitation arises, in part, due to
the waveform of the excitation signals currently employed in standard rheometric techniques. Single tone harmonic signals were established as the canonical tool of mechanical material characterization in the middle of the past century \cite{Menard1999}, and since then they have not been redesigned or optimized, as has been the case in other fields (e.g. acoustic measurements \cite{Fausti2000} or radar systems \cite{Klauder1960}), to achieve enhanced levels of resolution in both the time and frequency domain.

%

\begin{figure} [htb]
	\centering
	\includegraphics[width=\columnwidth]{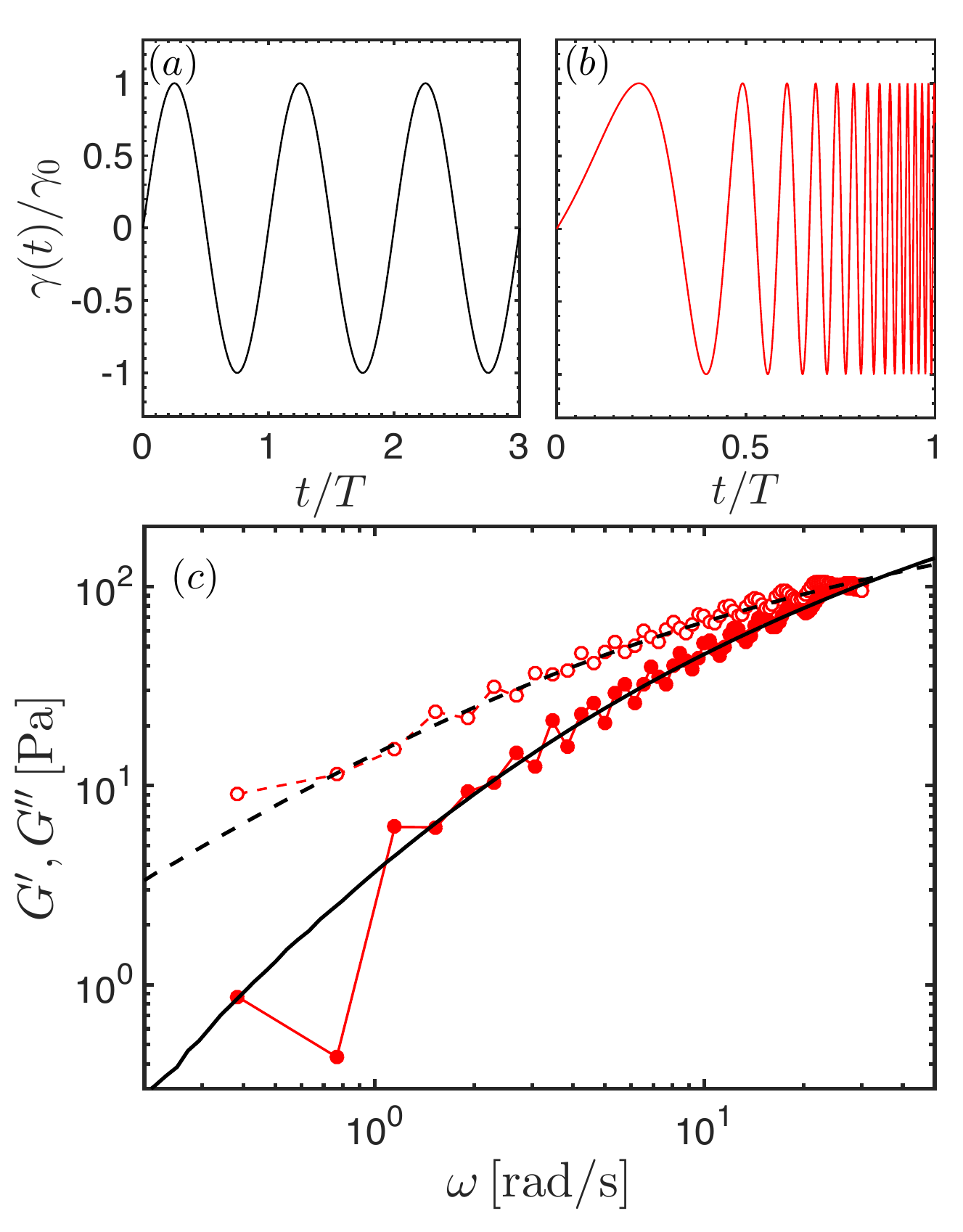}
	\caption{(a) Example of a single tone sine input: sinusoidal strain with constant amplitude $\gamma_0$ and frequency $\omega$; (b) Example of an exponential sine sweep or chirp: sinusoidal strain with constant amplitude and modulated frequency (see Eq.~(\ref{eq:chirp_x})); (c) Comparison of measured viscoelastic moduli for a reference PIB solution (see Appendix~\ref{app:materials} for details) obtained by a classical discrete frequency sweep ($G'$ \protect\Solidline, $G''$ \protect\dashedline)  with 30 points/decade (total experimental acquisition time greater than $30 \minutes$) and with application of only one chirp signal ($G'$ \protect\solidcircleA, $G''$  \protect\dashedcircleA) as described by Eq.~(\ref{eq:chirp_x}) with $T=14\s$, $\omega_1 = 0.3\rads$, $\omega_2 = 30\rads$, $\TB = 66$ at the same input strain $\gamma_0 = 6\%$.
		\label{fig:figure1}}%
\end{figure}

Standard test protocols for measuring the linear viscoelastic spectra of soft materials are typically based on periodic signals consisting of sine steps of constant frequency and amplitude (see Fig.~\ref{fig:figure1}a). These can either be combined sequentially as in frequency sweeps \cite{Kavanagh1998}, or they can be additively superposed for a discrete number of frequencies, such as in multi-wave techniques \cite{Holly1988}. Both of these methods present disadvantages when used with time-evolving or \emph{mutating} systems. A sequence of single tone inputs usually requires the longest total measuring time compared to other signals \cite{Muller2001},
even when limiting the number of frequencies to a small discrete set, as in discrete frequency sweeps \cite{Mours1994}. The additive superposition arising from multi-wave methods, on the other hand, can easily generate a signal with a total strain exceeding the linear viscoelastic limit of the test material, especially if many frequencies are summed together to increase the spectral content of the measured signal. Different types of signal that have been proposed to replace these standard ones include white noise \cite{Field1996} and step strain experiments, i.e., Heaviside-type excitations \cite{Tassieri2016}.
 
To produce harmonic test protocols with the time- and frequency-resolution necessary for studying mutating systems, we start from a different type of signal that has long been used for radar \cite{Klauder1960}, sonar \cite{Schock1989} and acoustic \cite{Fausti2000} measurements, called the sine sweep or \emph{chirp}. Chirps are frequency-modulated deterministic signals, i.e., their frequency is a continuously varying function of time (see for example Fig.~\ref{fig:figure1}b). Heyes \etal \cite{Heyes1994}, and more recently Ghiringhelli \etal \cite{Ghiringhelli2012}, Curtis \etal \cite{Curtis2014} and Rouyer and Poulesquen \cite{Rouyer2015} have used chirp signals to measure the viscoelastic spectra of different materials, claiming this method to be an Optimal Fourier Rheometry (OFR) technique \cite{Ghiringhelli2012,Curtis2014}.
However, a rigorous investigation of high resolution data on a model polymer network reveals that the measurement precision can be severely compromised especially when using short duration chirp signals. An example of this can be seen in Fig.~\ref{fig:figure1}c, in which we show the viscoelastic moduli of a semi-dilute PIB (poly(isobutylene)) solution (see composition in Appendix~\ref{app:materials}) measured with a standard frequency sweep (black lines) and with the OFR technique (red symbols). The moduli measured using a $14 \s$ OFR chirp signal are affected by significant local fluctuations that can compromise accurate determination of real material properties and computation of the linear viscoelastic spectrum. Such oscillations are a consequence of fluctuations in the power spectrum of the input signal, known as Fresnel ripples in signal processing \cite{Kowatsch1982}, and therefore they are not material-dependent features.

However, despite this issue chirps
have several features that are cardinal for the development of an optimal excitation signal for
mechanical spectroscopy. First, their crest factor, i.e. the ratio of the peak to the root mean square value of the waveform amplitude, is very low and comparable to that of a sine wave (1.45 instead of 1.414)  \cite{Muller2008,Pintelon2012}. This is very important because it allows us to constrain the amplitude of the signal within the linear range of the material response without sacrificing the corresponding spectral content (in contrast to multi-wave techniques \cite{Holly1988,Mours1994}). Second, as will become clearer in section~\ref{subsec:chirp_signal}, chirps can be designed to give a better signal-to-noise ratio (SNR) than white noise signals because their power spectral density is inversely proportional to the frequency
\cite{Muller2001,Muller2008,Pintelon2012}. This is essential in mechanical spectroscopy where force and torque measurements are corrupted by transducer noise primarily at lower frequencies. Finally, the waveform can be made periodic with both the initial and final value of the signal equal to zero \cite{Oppenheim1983,Pintelon2012}. This is another important aspect for mechanical measurements because it improves the numerical estimation of the signal power spectrum \cite{Pintelon2012}, but also because accumulated deformation or stress compounded from repeated tests could potentially affect the microstructure evolution and connectivity. 
%
As a consequence of these features, chirps have been widely used as excitation signals in a range of different applications \cite{Klauder1960,Schock1989,Fausti2000}, but they also commonly arise in nature from birdsong \cite{Stowell2014} to gravitational waves \cite{Abbott2016a} and, most notably, they are the signal forms used by both bats and dolphins for echolocation \cite{Au2007,Madsen2013}.

Inspired by these nearly optimal signals, we address the issues affecting the performance of chirps in mechanical spectroscopy by carefully designing their amplitude modulation to maximize the SNR while minimizing the unwanted spectral ripples. 
Combining experiments and detailed simulations on a model polymer solution, we show that it is possible to define an optimized chirp signal
that reduces the residual error in estimation of the linear viscoelastic spectrum by almost two orders of magnitude compared to a constant amplitude sine sweep. This \textit{Optimally Windowed Chirp}, or OWCh, allows us to determine the relaxation spectrum with essentially the same precision of the current discrete frequency sweep standard, while dramatically reducing the total test duration.
By applying the OWCh signal to a mutating protein gel undergoing gelation, we further show how time-resolved mechanical spectroscopy allows us to capture the evolution in the material viscoelastic properties of the gel in detail within a single experiment.


\section{Experimental Investigation}
\label{sec:exp_invetigation}

\subsection{Chirp signal construction}
\label{subsec:chirp_signal}


For a generic signal $x(t) = x_0\sin(\phi(t))$, the phase $\phi(t)$ is related to its instantaneous (angular) frequency by $\omega(t) = \di \phi(t)/\di t$, where $\omega(t)$ has units of $\rads$.
While classical sine waves maintain a constant frequency at any point in time, chirp signals are designed so that their instantaneous frequency is continuously changing, hence they are also commonly referred to as sine sweeps \cite{Muller2008,Isermann2011}. We are particularly interested in exponential chirps, i.e., sine sweeps with constant amplitude and frequency that increases exponentially in time following a relationship of the form:
	\begin{equation}
	\label{eq:chirp_freq}
	\omega(t) = \omega_2 \left( \frac{\omega_1}{\omega_2} \right)^{t/T} \, .
	\end{equation}
Here $T$ is the total length of the signal and $\omega_1$ and $\omega_2$ are respectively the initial  ($t=0$) and final ($t=T$) angular frequencies attained by the input signal. Integrating Eq.~(\ref{eq:chirp_freq}) and imposing that there be no initial phase shift ($\phi(t=0) = 0$),
we can derive an expression for the exponential chirp signal for any given values of $\omega_1$, $\omega_2$ and $T$:
	\begin{equation}
		\label{eq:chirp_x}
		x(t) = x_0 \sin \left\{ \frac{\omega_1 T}{\log(\omega_2/\omega_1)} \left[ \exp \left( \log(\omega_2/\omega_1) \frac{t}{T} \right) -1 \right] \right\} \, ,
	\end{equation}
A few previous studies have used this waveform $x(t)$ as a strain input $\gamma(t)$ to an unknown material (see Fig.~\ref{fig:figure1}b) from which one can obtain the output stress signal $\sigma(t)$ either experimentally, using a commercial rheometer \cite{Curtis2014,Rouyer2015}, or numerically, by integrating the appropriate set of equations of motion \cite{Heyes1994,Ghiringhelli2012}.  By taking the Discrete Fourier Transform (DFT) of $\gamma(t)$ and $\sigma(t)$ one can compute the complex modulus $G^\star(\omega)$ of the bulk material under investigation:
		\begin{equation}
	\label{eq:Gp_Gpp}
	G^\star (\omega) = \frac{\tilde{\sigma}(\omega)}{\tilde{\gamma}(\omega)}\,,
	\end{equation}
where $(\tilde{\cdot})$ indicates the Fourier transform. The elastic and loss moduli, $G'$ and $G''$, can subsequently be extracted as the real and imaginary part of the complex modulus, i.e., $G'(\omega) = \Re \{G^\star (\omega)\}$, $G''(\omega) = \Im \{G^\star (\omega)\}$.

More generally, if we consider the viscoelastic material being tested as a dynamical system with an unknown impulse response (commonly defined by the Boltzmann memory function $M(t)=\di G(t)/\di t$ in the time domain), then the process that we just described effectively corresponds to identification of the system's linear transfer function ($G^\star (\omega)$ in the frequency domain). Consequently, determining the most appropriate experimental excitation signal to achieve both time- and frequency-resolution in mechanical spectroscopy is equivalent to determining the fastest way to identify the transfer function of a mechanical system which, in the case of soft materials, is usually overdamped and characterized by a continuous power spectrum. 
Using nonlinear chirp signals, such as the exponential chirp in Eq.~(\ref{eq:chirp_x}), has advantages compared to other types of excitations especially in terms of the signal-to-noise ratio (SNR) achievable at low frequencies. In fact, exponential chirps have a pink frequency spectrum \cite{Muller2008}; that is, their power spectral density (per unit frequency) is inversely proportional to the frequency itself, leading to a constant power per unit octave (i.e., per fixed frequency interval $[\omega_1,\omega_2]$ such that $\omega_2 = 2 \omega_1, $). This feature of the chirp power spectral density is very important in all measurements where the noise floor dominates at lower frequencies and the possibility of exceeding the linear region of the system is of concern. An exponential chirp sequence is thus particularly well suited to mechanical spectroscopy of soft materials (as well as airborne and acoustical measurements) because it allows us to maintain sufficient SNR even at small $\omega$ without dangerously increasing the power input \cite{Muller2008} and hence exceeding the linear viscoealstic limit of the material. 

It can be seen from Eq.~(\ref{eq:Gp_Gpp}) that, when using a broadband excitation signal, consistent determination of the frequency response of a system directly depends upon the accurate determination of the power spectral density of both the input and output signals. These, in turn, are known to greatly depend on the Time-Bandwidth constant of the signal ($\TB$), defined as the product of the length of the chirp to its nominal bandwidth, such that \mbox{$\TB = T(\omega_2 - \omega_1)/(2\pi)$} \cite{Kowatsch1982,Misaridis2005a}.
The main artifact that decreases the accuracy of the spectral estimation is related to the appearance of side lobes in the chirp power spectrum. These ripples
result from the DFT of the inherent rectangular envelope of the chirp defined in Eq.~(\ref{eq:chirp_x}) \cite{Pintelon2012,Oppenheim1983}. In fact, imposing a chirp with constant amplitude mathematically corresponds to multiplying the frequency modulated signal with a normalized square wave of length $T$. When one takes the DFT of this product, the result is given by the convolution of the two Fourier transforms and therefore the side lobes present in the spectrum of the envelope produce similar ripples in the frequency spectrum of the chirp.
Another source of spectral leakage
is related to the absence of periodicity that arises if $x(0)\neq x(T)$ \cite{Pintelon2012} (or, equivalently, if the signal does not contain an integer number of periods), a condition experimentally very hard to achieve, but that can in theory be avoided by carefully adjusting the length of the signal.

The magnitude of these ripples in the Fourier domain can be decreased by designing chirps with large values of the time-bandwidth constant, i.e., $\TB \gg 100$ \cite{Misaridis2005b}.
Since the frequency range accessible in mechanical instruments is limited, working with large values of $\TB$ implies using longer chirps and therefore compromising the time (or frequency) resolution for a fast mutating system. On the other hand, working
with shorter signals ($\TB < 100$) usually yields significant oscillations in the viscoelastic spectra as shown in
Fig~\ref{fig:figure1}c for the reference PIB solution.
The discrete (red) symbols are the results obtained from Eq.~(\ref{eq:Gp_Gpp}), using the input strain chirp in Eq.~(\ref{eq:chirp_x}) with $\omega_1 = 0.3 \rads$, $\omega_2 = 30 \rads$, $\gamma_0 = 0.06$ and $T=14\s$. The corresponding time-bandwidth constant is $\TB \simeq 66$.
Using a single chirp signal we can reduce the experimental test duration required to measure the viscoelastic response of the material by almost two orders of magnitude when compared to classical frequency sweeps (solid and dashed black lines in Figure~\ref{fig:figure1}c), but the estimates of the viscoelastic moduli thus obtained are substantially affected by spectral leakage, as the oscillations observed in both $G'$ and $G''$ in Figure~\ref{fig:figure1}c clearly show.
%

In order to develop a truly optimized excitation signal for mechanical spectroscopy of soft materials, we therefore need to ensure that chirp sequences can be used even when the experimental conditions are not ideal, especially for short input durations, or small time-bandwidth constants. The natural time-scale that sets how short the input signal should be depends on the characteristic mutation time of the material of interest $\tau_{mu} = (\di \ln g/ \di t)^{-1}$ (where $g$ is any property of interest, e.g., $G^\star$) \cite{Mours1994}. We can construct a dimensionless number given by the ratio of the signal length to the characteristic mutation time scale of the material itself, i.e., a mutation number $N_{mu} = T/\tau_{mu}$ \cite{Winter1988}. To have a reliable time-resolution we require $N_{mu} \ll 1$ and a conservative value for the critical mutation number that guarantees such a condition, at least for gelation processes, appears to be $N_{mu}<N_{mu,cr} = 0.15$ \cite{Mours1994,Curtis2014}. This sets the maximum length of the input signal for any given value of $\tau_{mu}$.

\subsection{Windowed Chirp}
\label{subsec:windowed_chirp}

Several signal processing techniques are known to improve the accuracy of the spectral estimates of different input signals, the most common being pre-whitening and tapering \cite{Priestley1981}. Their effectiveness, however, strongly depends on the specific application for which the signal is used and the type of frequency response expected.
Since our main objective is to reduce spectral leakage and ultimately obtain a smooth transfer function for the material in Fourier space, there are two main issues that must be tackled: the presence of side lobes projected from the square envelope defining the amplitude of the chirp
and any additional spectral leakage due to the absence of periodicity.

To address both these concerns we take inspiration from sonar signals that have been optimized and improved over thousands of years: specifically, the biosonar signals used by bats and dolphins for echolocation \cite{Au2007,Madsen2013}. Most of the pressure waves emitted by these animals are frequency modulated signals, remarkably close to logarithmic sweeps, however they also feature an amplitude modulation that depends on the specific target \cite{Jones2005}. From a mathematical point of view, amplitude modulation can be achieved by using an appropriate function to prescribe the envelope of the excitation signal, commonly referred to as a window function $w(t)$ \cite{Priestley1981}.
Windowing is widely used in signal processing as a way to select and analyze a portion of a longer recorded signal, the main advantage being that the specific function can be designed to smoothly vary between zero (outside of the window) and unity (usually at the center of the window). This restores a general periodicity which reduces part of the spectral leakage and also allows control of the magnitude of the side-lobes in the spectral domain (known as side-lobe level).
In its original form (Eq.~(\ref{eq:chirp_x})) the chirp signal has an envelope defined by a square wave known as a rectangular window, or Dirichlet window, $w(t) = 1$ for $0 \le t \le T$. This is a discontinuous function and, as a consequence, it is characterized by the highest side-lobe level when compared to other window functions \cite{Harris1978}.

Choosing the appropriate window is a matter that has generated a considerable body of work, since different applications have different requirements in terms of window performance \cite{Harris1978,Priestley1981}.
Here, we propose to modify the original frequency-modulated, constant-amplitude chirp by using a particular cosine-tapered function (also called a Tukey window \cite{Tukey1968}) defined as:
%
%
\begin{equation}
\label{eq:window}
{w(t)} = 
\\
\begin{dcases}
 \cos^2 \left[ \frac{\pi}{r} \left( \frac{t}{T} - \frac{r}{2}\right) \right],&\frac{t}{T} \le \frac{r}{2}\\
1,& \frac{r}{2} < \frac{t}{T} < 1 - \frac{r}{2} \\
\cos^2 \left[ \frac{\pi}{r} \left( \frac{t}{T} - 1 + \frac{r}{2}\right) \right],& \frac{t}{T} \ge 1- \frac{r}{2}
\end{dcases}
\end{equation}

%
where $r$ is a dimensionless tapering parameter that allows us to tune how rapidly the amplitude is modulated within the length of the signal $T$.
\vspace{0.1cm}

This window has two interesting limits: for $r=0$ it effectively becomes a rectangular window, while for $r=1$ it corresponds to the well known Hann window that is characterized by $\mathcal{C}^1$ continuity \cite{Harris1978}. Values of $r$ greater than unity are equivalent to using a Hann window with a lower maximum amplitude. For $r>0$, the Tukey window has smaller sidelobe levels in the Fourier domain than the rectangular window, and the functional form also guarantees that the signal is always zero at both ends, thus ensuring periodicity while introducing side ripples with smaller amplitude. Additionally, although with this choice we have fixed the shape of the tapering function, we also maintain an important degree of freedom thanks to the presence of the parameter $r$, that allows us to explore effects of the extent of tapering on both the input and output signal power spectrum.
Additional discussion about the importance of the specific window shape is given in Appendix~\ref{app:error_analysis}.

\begin{figure*}[htb]
	\centering
	\includegraphics[width= 0.8\textwidth]{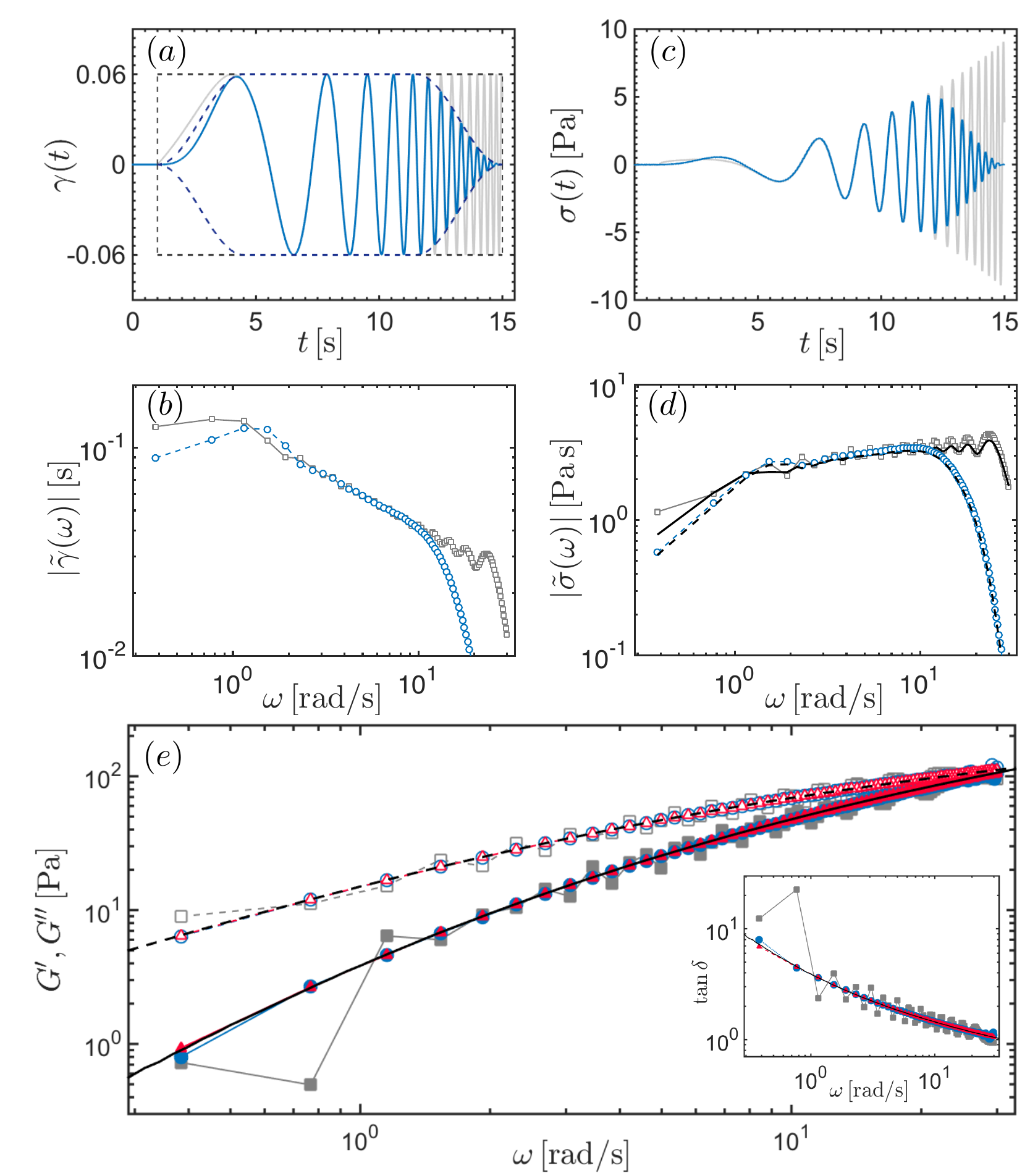}
	\caption{Data presented is from experiments on the reference PIB solution. All chirp signals have the following parameters $\omega_1 = 0.3 \rads$, $\omega_2 = 30 \rads$, $\gamma_0 = 0.06$, $T=14\s$ ($\TB \simeq 66$) but different degree of tapering, i.e. different values of $r$. (a) Sine sweep in strain $\gamma(t)$ (as proposed by Ghiringhelli et al \cite{Ghiringhelli2012}) with constant signal amplitude $r=0$ (\protect\SolidlineB). The enclosing dashed line (\protect\dashedlineB) emphasizes the square nature of the signal envelope. Superposed is a windowed chirp with $r=50\%$ (\protect\SolidlineC) together with its tapered envelope (\protect\dashedlineC). Both signals begin at time $t_w=1\s$ as described in the text; (b) Single-sided amplitude spectrum of the strain signal for the same values of the tapering parameter $r$ as in part (a): $r=0$ (\protect\solidsquareDw) and $r=50\%$ (\protect\dashedcircleD); (c) Measured stress response as a function of time corresponding to the input signals of part (a): $r=0$ (\protect\SolidlineB), $r=50\%$ (\protect\SolidlineC); (d) Single-sided amplitude spectrum of the stress signals shown in part (c). Data from the measured stress ($r=0$ \protect\solidsquareDw, $r=50\%$ \protect\dashedcircleD) are compared to the error-free spectrum ($r=0$ \protect\Solidline, $r=50\%$ \protect\dashedline) obtained in the frequency domain by multiplying the theoretical transfer function from the fractional Maxwell model, i.e. the complex modulus $G^\star(\omega)$ of the material, and the DFT of the experimental strain input; (e) Comparison of the viscoelastic moduli $G^\star (\omega) = G'(\omega) + i G'' (\omega)$ obtained by a classical frequency sweep  ($G'$ \protect\Solidline, $G''$ \protect\dashedline) with the results obtained from a windowed chirp with different tapering parameters: $r=0$ ($G'$ \protect\solidsquareD, $G''$ \protect\dashedsquareD),$r=10\%$ ($G'$ \protect\solidtriangleE, $G''$ \protect\dashedtriangleE) and $r=50\%$ ($G'$ \protect\solidcircleD, $G''$ \protect\dashedcircleD). Inset shows $\tan \delta (\omega) = G''(\omega)/G'(\omega)$.
		\label{fig:figure2}}%
\end{figure*}

If we apply the Tukey window in Eq.~(\ref{eq:window}) to the original chirp  signal introduced in Eq.~(\ref{eq:chirp_x}), we obtain a \emph{windowed chirp} that has the following form:
%
\begin{widetext}
	\begin{equation}
	\label{eq:windowed_chirp}
	x(t) = x_0
	\begin{dcases}
	\cos^2 \left[ \frac{\pi}{r} \left( \frac{t}{T} - \frac{r}{2}\right) \right]\sin \left\{ \frac{\omega_1 T}{\log(\omega_2/\omega_1)} \left[ \exp \left( \log(\omega_2/\omega_1) \frac{t}{T} \right) -1 \right] \right\}\,, &\frac{t}{T} \le \frac{r}{2}\\
	\sin \left\{ \frac{\omega_1 T}{\log(\omega_2/\omega_1)} \left[ \exp \left( \log(\omega_2/\omega_1) \frac{t}{T} \right) -1 \right] \right\}\,,  & \frac{r}{2} < \frac{t}{T} < 1 - \frac{r}{2} \\
	\cos^2 \left[ \frac{\pi}{r} \left( \frac{t}{T} - 1 +\frac{r}{2}\right) \right]\sin \left\{ \frac{\omega_1 T}{\log(\omega_2/\omega_1)} \left[ \exp \left( \log(\omega_2/\omega_1) \frac{t}{T} \right) -1 \right] \right\}\,, & \frac{t}{T} \ge 1- \frac{r}{2}
	\end{dcases}
	\end{equation}
\end{widetext}
Equation~(\ref{eq:windowed_chirp}) can be implemented into the arbitrary wave function of a commercial strain-controlled rheometer (ARES-G2, TA Instruments) and the strain signal directly generated by the motor using such a command with a digital  sampling frequency $f_s = 500 \Hz$ is plotted in Fig.~\ref{fig:figure2}(a) for
$r=50\%$ (solid blue line) and for $r=0$, i.e., constant amplitude, (solid gray line). All of the other parameters characterizing the chirp signal are the same as in Fig.~\ref{fig:figure1}b, namely $\omega_1 = 0.3 \rads$, $\omega_2 = 30 \rads$, $\gamma_0 = 0.06$ and $T=14\s$ ($\TB \simeq 66$). A waiting time of $t_w = 1 \s$ is also imposed before applying any chirp input; this is only necessary from a signal conditioning perspective in order to guarantee the strain output from the sensor is zero at the beginning of the signal and does not have any implication from a theoretical standpoint.
The shape of the envelope of each signal is also shown, marked by a dashed line with the same color code. From this plot we can see more directly how the Tukey window guarantees that the chirp signal is zero both at the beginning and at the end, while smoothly modulating its amplitude from zero to a fixed maximum value which is here set to be $\gamma_0=0.06$.

The effect of this amplitude modulation on the Fourier transform of the chirp signal is highlighted in Fig.~\ref{fig:figure2}(b), where the amplitude spectrum (i.e., the square root of the power spectrum) of both the signals plotted in Fig.~\ref{fig:figure2}(a) are shown (same color code applies). While the spectrum of the chirp with $w(t) = 1$ is strongly affected by ringing across the entire frequency range, the windowed chirp has a much smoother transform with no visible ringing artifacts. However, it is important to notice also that while the amplitude spectrum for $r=0$ oscillates around an ideal pink spectral response (with $|\tilde{\gamma}(\omega)| \sim \omega^{1/2}$) across the entire signal bandwidth, as soon as $r>0$, a portion of the power spectrum is affected by the reduced amplitude of the signal. This reduction in signal strength influences the lowest and highest frequencies excited during the window rise and fall times respectively. As a consequence, windowing affects an increasingly wider range of frequencies for large values of the tapering parameter, yielding inherently different power spectra for each value of $r$.


The measured shear stress response $\sigma(t)$ of the reference PIB solution to the strain input signals shown in Fig.~\ref{fig:figure2}(a) are plotted in Fig.~\ref{fig:figure2}(c) with their corresponding amplitude spectra in Fig.~\ref{fig:figure2}(d).
Similarly to the strain input signal, the shear stress measured as a response of the windowed chirp is also forced to be zero at both the beginning and at the end of the time interval $T$. By contrast the stress measured as a response to the chirp with $r=0$ is clearly non-zero at the end of the signal. The corresponding amplitude spectra computed directly from the DFT of $\sigma(t)$ and plotted in Fig.~\ref{fig:figure2}(d) also show the presence of side lobes as observed in the strain input.

From the DFT of the strain and stress signals we can also compute the elastic and loss moduli via Eq.~(\ref{eq:Gp_Gpp}), following the procedure detailed in Appendix~\ref{app:materials}. The results are shown in Fig.~\ref{fig:figure2}(e). The black lines indicate reference values measured using a classical frequency sweep (solid line is $G'(\omega)$, dashed line is $G''(\omega)$), while the symbols represent the results from different windowed chirps (closed symbols for $G'(\omega)$, open symbols for $G''(\omega)$). This figure clearly highlights how the suppression of side lobes in the Fourier transform of both output (stress) and input (strain) signals also dramatically reduces the oscillations in the computed moduli with respect to the case of a non-windowed chirp. The inset shows the same data presented in terms of the phase angle $\tan \delta = G''/G'$, which is an important parameter to follow when studying gelation  \cite{Winter1986,Chambon1987} as we explain in more detail in section~\ref{sec:casein}. This representation is also more sensitive to the presence of ripples in the power spectrum as the ratio of two quantities can magnify the influence of oscillations. Visual inspection of each signal shows that the application of an amplitude modulation is extremely efficient in suppressing spectral leakage, as comparison with results from the classical frequency sweep (solid black line) shows. However, we can also see that there remain differences between the values obtained with $r=10\%$ and $r=50\%$, suggesting that not all values of the tapering parameter give the same level of enhancement. It is therefore compelling to explore how to determine the value of $r$ that provides the best estimate of the viscoelastic spectrum.

\subsection{Optimization of windowed chirps}
\label{subsec:optimization}

\begin{figure} [htb]
	\centering
	\includegraphics[width=\columnwidth]{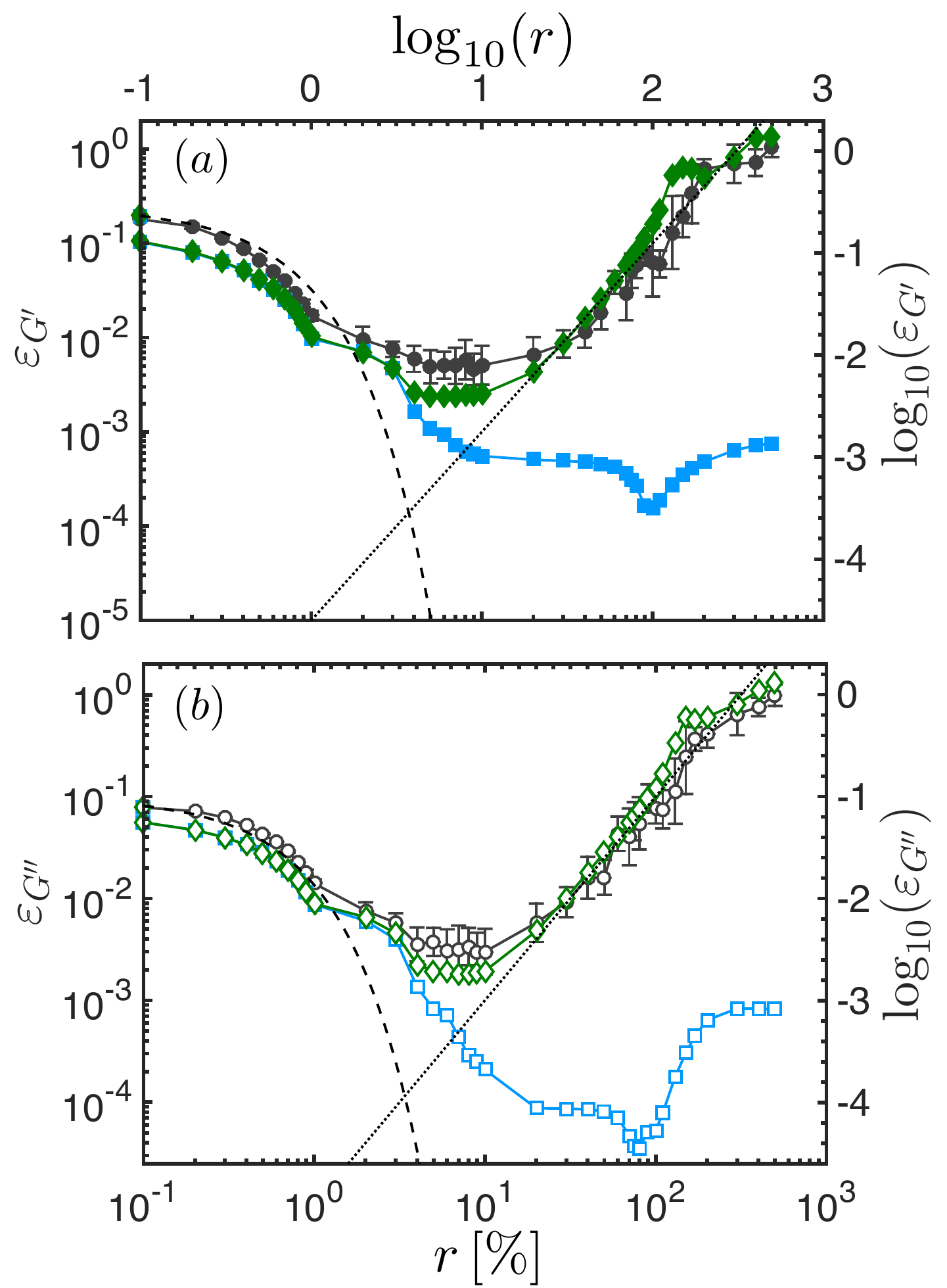}
	\caption{Summary and comparison of experiments and numerical simulations on the reference PIB solution. The plots show the RMS error in the determined storage modulus (a) and loss modulus (b), defined in Eqs.~(\ref{eq:error_gp})~and~(\ref{eq:error_gpp}), as a function of the tapering parameter $r$. Circles ($\varepsilon_{G'}$ \protect\solidcircleF, $\varepsilon_{G''}$ \protect\solidcircleFw) show the experimental results averaged over 6 different realizations. Error bars correspond to one standard deviation. Squares ($\varepsilon_{G'}$ \protect\solidsquareF, $\varepsilon_{G''}$ \protect\solidsquareFw) are the results from numerical simulations without noise, while diamonds ($\varepsilon_{G'}$ \protect\soliddiamondE, $\varepsilon_{G''}$ \protect\soliddiamondEw) show the trend from simulations when Gaussian noise is added to the stress signal in order to mimic experimental conditions.
		\label{fig:figure3}}%
\end{figure}

In order to determine the optimal value of the tapering parameter, we conducted a systematic series of experiments on the PIB solution, testing windowed chirps with increasing values of $r$ in the range $0 \le r \le 5$. Each sequence of tests was preceded and followed by a frequency sweep at the same amplitude with respectively $5$ and $30$ data points per decade. These sweeps were used to ensure that there was no significant change in the viscoelastic spectrum of the reference material during the tests. After post-processing the data collected from the rheometer (see Appendix~\ref{app:materials}), we computed the storage and loss moduli using Eq.~(\ref{eq:Gp_Gpp}). In order to quantify the differences between the moduli measured with the classical discrete frequency sweep and those determined with the windowed chirps, we then computed the root mean square (RMS) error between the two estimates of the relaxation spectrum over all the frequencies available:
	\begin{align}
	\label{eq:error_gp}
	\varepsilon_{G'} (r) = \underset{\omega_i}{\mathrm{RMS}}\left\{\log \left[ \frac{G'_{chirp}(\omega_i; r)}{G'_{Sweep}(\omega_i)}\right]\right\} \, , \\
	\label{eq:error_gpp}
	\varepsilon_{G''} (r) = \underset{\omega_i}{\mathrm{RMS}}\left\{\log \left[ \frac{G''_{chirp}(\omega_i; r)}{G''_{Sweep}(\omega_i)}\right]\right\} \, ,
	\end{align}
thus obtaining a single value of the error for each contribution to the complex modulus and for each value of $r$ in one series of tests. To match the frequencies at which the moduli are computed when using a chirp signal with those of the frequency sweep, we performed a cubic spline interpolation of the discrete frequency sweep data set that was obtained with $30$ points per decade. The experiments were repeated at least $6$ times and the values of the error obtained for each $r$, averaged over all the experiments, are shown with dark gray circles in Fig.~\ref{fig:figure3}(a) and (b) for the elastic and loss moduli respectively. The error-bars represent one standard deviation.

The observed trends are robust for both contributions to the complex modulus: there is an initial region for small values of the tapering parameter ($r \le 1\%$) that shows an approximately exponential decay of the form $\varepsilon \sim \exp(-200r)$ as evidenced by the dashed black line. For larger values of $r$ the decrease is less pronounced until the error reaches an intermediate minimum plateau. Finally, for $r > 15 \%$ the error begins to increase quadratically, as highlighted by the dotted black line. These tests thus show that there is a range of values of the tapering parameter, for $r \in [6, 15] \%$, that minimizes the error obtained in the viscoelastic spectrum measured with a windowed chirp, and reduces it by almost two orders of magnitude with respect to the case of a non-windowed chirp.
The magnitude of this improvement brings the difference between results for a classical frequency sweep and a windowed chirp  within the range of experimental accuracy that can be obtained with single tone frequency sweeps. We therefore conclude that
an Optimally Windowed Chirp (OWCh) can indeed be used to measure the linear viscoelastic response of any soft material with the same level of accuracy as a classical frequency sweep \cite{Schultheisz2000}, while reducing by several orders of magnitude the total test duration required. An analogous optimization procedure was also performed on a micellar solution and the corresponding results were in good agreement with the results for the PIB solution (see Appendix~\ref{app:micellar_solution} for additional discussion).

While this experimental investigation alone is sufficient for determining the optimal range of values for the tapering parameter, it does not provide sufficient insights to understand why we observe this trend in the estimation error.
Having a theoretical framework or numerical tool to perform the same type of tests and compare the results to experimental data offers the additional benefit of gaining new insights, and understanding in more depth, the performance of windowed chirp signals within the specific context of mechanical spectroscopy.

\section{Numerical Investigation}
\label{sec:numerical_simulations}

\subsection{Constitutive equation for PIB solution}
\label{subsec:PIBconst_eq}

The first step in performing any analytical or numerical analysis of our OWCh protocol is to determine an appropriate constitutive equation for the PIB solution. The dependence of $G'$ and $G''$ on $\omega$, as first shown in Fig.~\ref{fig:figure1}c, is characteristic of a viscoelastic fluid; at low frequencies the loss modulus is higher than the elastic storage modulus and therefore the material is expected to flow for any non-zero applied stress (with $\lim \{\tan \delta\} \rightarrow \infty$ for $\omega \rightarrow 0$). The simplest viscoelastic model that captures such a behavior is the Maxwell model and its mechanical analog consists of a dashpot of viscosity $\eta$ in series with a spring of shear modulus $G$~\cite{Bird1987}. The Maxwell model is sufficient for simple viscoelastic liquids characterized by a single time-scale that corresponds to the relaxation time of the material $\tau = \eta/G = 1/\omega_c$. The model predicts that the storage and loss moduli should increase, respectively, quadratically and linearly at low frequencies ($\omega \ll 1/\tau$). However very few viscoelastic liquids can be quantitatively described by the Maxwell model with the exception of micellar solutions (see Appendix~\ref{app:micellar_solution} for an example). The majority of real viscoelastic fluids, including the PIB solution used in this study, have a polydisperse microstructure and a broad relaxation spectrum governed by Rouse-Zimm dynamics. This is reflected in the frequency response, which features different power-law regimes than those predicted by the Maxwell model. In order to capture the presence of multiple relaxation modes one can use a generalized Maxwell model, obtained by combining several Maxwell models in parallel, but this increases the number of fitting parameters substantially and does not convey any additional microstructural insight \cite{Tschoegl1989}.
More recently, several researchers have considered the Fractional Maxwell Model (FMM) \cite{Koeller1984}, which consists of two fractional or ``spring-pot'' elements arranged in series, to capture these multiple relaxation modes using a very small number of fitting parameters.
A fractional element, first introduced in the context of viscoelasticity by Scott-Blair \cite{Scott-Blair1944}, is a mechanical element characterized by the following constitutive equation:
	\begin{equation}
		\label{eq:spring-pot}
		\sigma(t) = \mathbb{V} \frac{\di^\alpha}{\di t ^\alpha} \gamma(t) \, ,
	\end{equation}
where $0 \le \alpha \le1$ and the parameter $\mathbb{V}$, which sets the scale of the stress, is sometimes called a quasi-property since its units are $\Pas^\alpha$ \cite{Jaishankar2012}. The fractional derivative represented in Eq.~(\ref{eq:spring-pot}) has a precise mathematical definition and can therefore be appropriately computed, e.g., using a Caputo derivative \cite{Caputo1995}. As explained in more detail in \cite{Jaishankar2012}, a spring-pot element has both an elastic and a viscous nature, the balance of which depends on the value of $\alpha$. In fact, in the limit of $\alpha =0$ one recovers the classical elastic response of a Hookean spring (with $\mathbb{V} \rightarrow G$), while for $\alpha = 1$, Eq.~(\ref{eq:spring-pot}) is equivalent to the constitutive equation for a Newtonian fluid (with $\mathbb{V} \rightarrow \eta$). For this reason the mechanical analog of a fractional element is also called a spring-pot. If we indicate the quasi-properties of the two elements in the model with $(\mathbb{V},\alpha)$ and $(\mathbb{G},\beta)$, then the constitutive equation for the FMM can be written as \cite{Jaishankar2012}:
	\begin{equation}
	\label{eq:FMM}
		\frac{\di^{\alpha-\beta}}{\di t^{\alpha-\beta}} \sigma (t) = \mathbb{G} \frac{\di^\alpha}{\di t^\alpha} \gamma(t) - \frac{\mathbb{G}}{\mathbb{V}} \sigma(t)
	\end{equation}
where we assume $\alpha > \beta$ without loss of generality. By taking the Fourier transform of Eq.~(\ref{eq:FMM}) one can easily obtain an analytic expression for the complex modulus $G^\star (\omega)$ of the fractional Maxwell model (i.e. its transfer function) \cite[see][for more details]{Jaishankar2014}. The expressions for $G'(\omega)$ and $G''(\omega)$ can easily be fitted to the experimental data to determine the values of the four constitutive parameters. For the PIB solution, the fitting procedure yields the following values: $\alpha = 1$, $\mathbb{V} = \eta = 18 \Pas$, $\beta = 0.36$ and $\mathbb{G} = 50 \,\mathrm{Pa\, s^{0.36}}$. Interestingly, one of the elements is effectively a classical viscous dashpot, indicating that the terminal response corresponds to a viscous liquid with $\lim {G'' (\omega)} \rightarrow \eta \omega$ for $\omega \rightarrow 0$. This also simplifies Eq.~(\ref{eq:FMM}) considerably from the point of view of numerical integration procedures. In fact, analyzing the behavior of a FMM analytically for a windowed chirp strain input is quite complex and does not provide a closed form expression that can be computed without any numerical assistance. Therefore, it is actually simpler to study the system response by directly integrating Eq.~(\ref{eq:FMM}) with a strain input given by Eq.~(\ref{eq:windowed_chirp}).

\subsection{Numerical simulations of Windowed Chirps}
\label{subsec:simulations_FMM}


To better understand trends in the error as a function of the tapering parameter, we performed a series of simulations using the fractional Maxwell model. Simulations were setup to accurately integrate Eq.~(\ref{eq:FMM}) so that the numerical error would be smaller than the error due to spectral leakage. To this end, we used a generalized backward differentiation formula with a time step $h = 2 \cdot 10^{-6} \s$, as determined by comparing results with increasingly smaller time steps.
To reduce truncation errors and thanks to the fact that $\alpha =1$, the analytical expression for the strain rate $\dot{\gamma}(t)$ was directly implemented as input instead of the strain $\gamma(t)$. The resulting stress signal was calculated for the same values of the tapering parameter used in the experiments and then down-sampled by cubic spline interpolation to obtain a signal with the same sampling frequency used in the experiments ($f_s = 500 \Hz$). Each data set of strain and stress thus obtained was processed with the procedure used for experimental data (see Appendix~\ref{app:materials}) to obtain values of $G'$ and $G''$. Analogously to experiments, an average error for each viscoelastic modulus was obtained using Equations~(\ref{eq:error_gp}) and~(\ref{eq:error_gpp}), where the values corresponding to a frequency sweep were directly calculated from the analytical expressions for $G^\star (\omega)$ (see \cite{Jaishankar2014}).
	
The results of this analysis are shown with light blue squares in Fig.~\ref{fig:figure3}(a) and (b) for the storage and loss moduli respectively. The trends closely approximate the experimental ones for values of the tapering parameter up to about $6\%$, that is, until the beginning of the optimal range of $r$. However, while the experiments show a plateau in the magnitude of the RMS error followed by a subsequent quadratic increase, numerical simulations display a continuously decreasing error in both $G'$ and $G''$, with a minimum being obtained for $r=100\%$, corresponding to a full Hann window. This is consistent with the observation that the Hann window has lower side lobes than any of the other tapered cosine windows ($0 \le r < 1$) while retaining the benefit of general periodicity (thanks to the signal having the same zero value at both ends).
This can be further appreciated by comparing the amplitude spectrum of the stress signal as computed by taking the DFT of the measured data versus the ``ideal'' spectrum computed as $|G^\star (\omega) \tilde{\gamma}(\omega)|$, where $G^\star (\omega)$ is the complex modulus analytically derived for the FMM. As Fig.~\ref{fig:figure2}(d) highlights, the decrease in spectral leakage for increasing $r$ is evident from the improvement of the estimate of the ideal spectrum (solid and dashed lines).
If there were no extrinsic factors to account for when considering the optimization problem, our numerical simulations therefore show that the best window would be a Tukey window with $r=100\%$ (or Hann window).

Nonetheless, the experiments clearly show that this is not the case when performing measurements with real rheometric instrumentation and this should not come as a surprise. In fact,
while numerical simulations are free of any noise in either the strain input or in the stress output signal (the numerical integration errors in the stress signal being small enough not to interfere with the error analysis pursued here), the recorded experimental signals are not noise free. In particular, the shear stress signal measured by the force rebalance transducer of the rheometer has its own transfer function and frequency characteristics. Even when left unperturbed, small ambient vibrations are detected by the sensor transducer and this is clearly evident especially when looking at signal traces acquired without any applied deformation (that is, imposing zero strain).

To test this hypothesis, we analyzed the stress signals obtained from measurements performed in the initial waiting time during which the strain is maintained at zero. Taking the root mean square error of the signal for $0 \le t \le t_w$, we determined a noise level close to $0.03 \Pa$. We therefore added a Gaussian white noise with this standard deviation to the stress signals computed numerically and repeated the post-processing protocol to obtain the new estimation of the viscoelastic moduli and their associated errors. The results, obtained using the same noise vector for all stress signals, are shown in Fig.~\ref{fig:figure3}(a) and (b) with (green) squares. The trends are now much closer to the ones observed experimentally, thus confirming that the minimum plateau value of the RMS error and the subsequent power-law increase for large values of the tapering parameter are both due to the presence of a non-negligible noise floor in the shear stress transducer (which decreases the SNR of the output signal). This analysis explains why, in the presence of noise, the optimal window corresponds to an intermediate value of the tapering parameter. Although spectral leakage is decreased, as $r$ increases the power spectrum of the input signal also decreases for a progressively wider range of frequencies at both ends of the spectrum, thus deteriorating the SNR of the output signal.
Additional considerations regarding the trends in the RMS error and the choice of the tapering function form are given in Appendix~\ref{app:error_analysis}.


So far we have used viscoelastic fluids which are inherently stable over time (with $\tau_{mu} \rightarrow \infty$), and thus constant material properties, in order to optimize the experimental procedure. In the next section we proceed to show how an optimally windowed chirp can be used to monitor the viscoelastic properties of a mutating system by looking at the gelation dynamics of an acid-induced protein gel \cite{Bremer1990,Totosaus2002,Leocmach2014,Keshavarz2017}.

\section{Application to a mutating system}
\label{sec:casein}


Time-resolved mechanical spectroscopy is crucial when studying systems such as gels undergoing a liquid-to-solid transition. 
The principal feature characterizing a liquid-to-solid transition is  knowledge of the Gel Point (GP), i.e., the point in time when the material microstructure percolates to form a sample-spanning network. In many systems, the self-similarity of the gel structure results in a characteristic broad distribution of relaxation modes that is reflected in a power-law dependence of both viscoelastic moduli on the angular frequency, such that $G' \sim \omega^\alpha$ and $G'' \sim \omega^\alpha$ where $0<\alpha<1$ \cite{Winter1997}. As a consequence, their ratio $\tan \delta = G''/G'$, or equivalently the phase angle $\delta$ between the stress and the strain signal, is effectively independent of the angular frequency itself \cite{Winter1986,Chambon1987,Winter1997}. A gel that shows this type of viscoelastic response is also referred to as a \emph{critical gel}.
%
%
%
%
Being able to accurately resolve the GP is therefore strongly dependent on the ability to have enough frequency resolution in the linear viscoelastic power spectrum at any point in time, while still maintaining a small mutation number.

\begin{figure*}
	\centering
	\includegraphics[width=\textwidth,trim={0 1.8cm 0 1.8cm}]{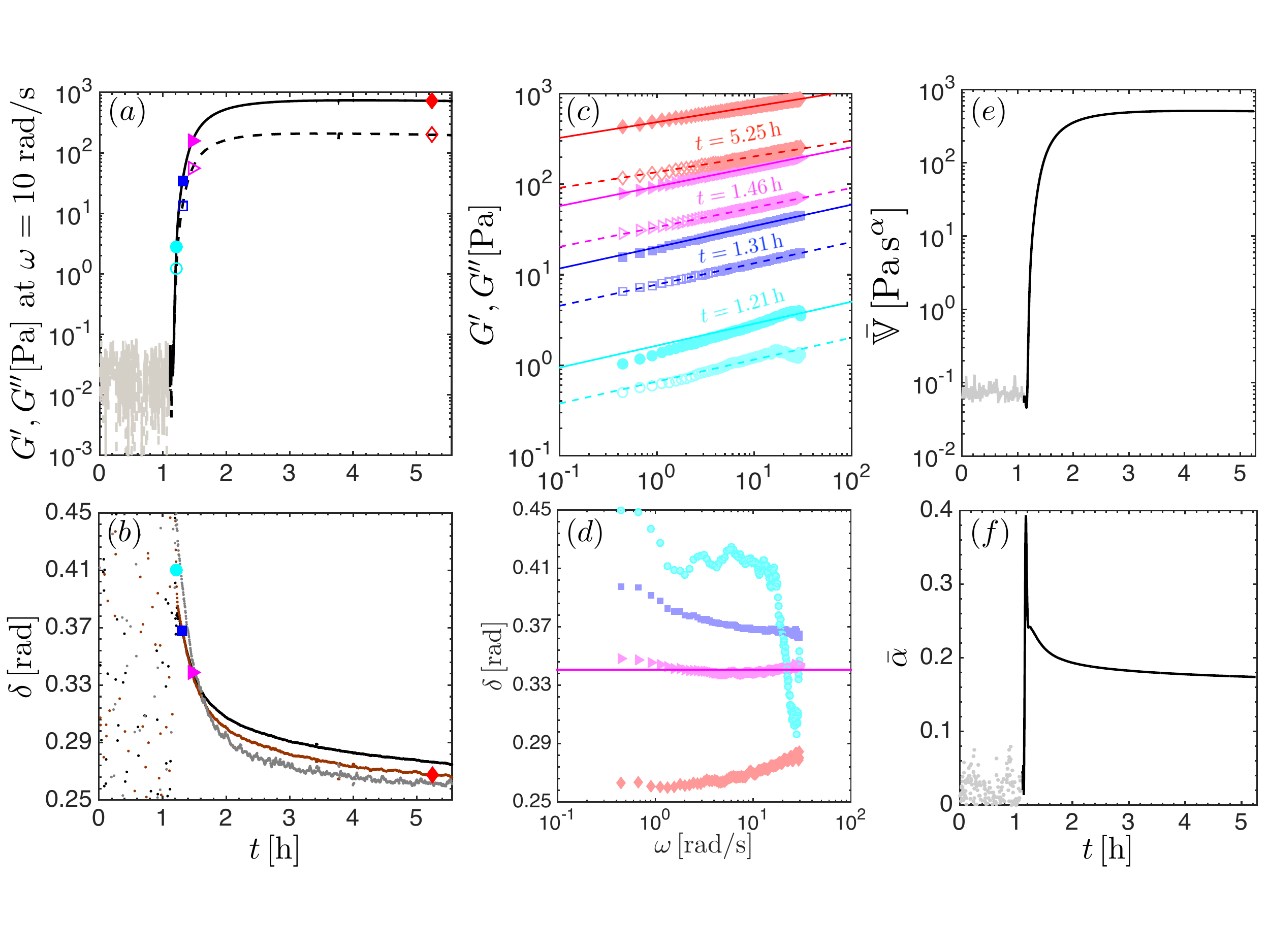}
	\caption{Time- and frequency-resolved gelation of a casein gel obtained using an OWCh sequence with $r=10\%$, $T=14\s$, $\omega_1 = 0.3\rads$, $\omega_2 = 30\rads$, $\TB = 66$, $t_w = 1\s$ and $\gamma_0 = 1\%$. (a) Evolution of the viscoelastic moduli ($G'$ \protect\Solidline, $G''$ \protect\dashedline) as a function of gelation time for one selected test frequency ($\omega = 10\rads$); (b) Phase angle $\delta$ as a function of time for different frequencies: $\omega = 0.3 \rads$ (\protect\SolidlineB), $\omega = 10 \rads$ (\protect\SolidlineG) and $\omega = 20 \rads$ (\protect\Solidline). At the gel point (indicated by \protect\triangleG) all three frequencies pass through a single value of $\delta$ because the phase angle is independent of frequencies \cite{Winter1986,Chambon1987,Winter1997}; (c) Viscoelastic moduli ($G'$ closed symbols, $G''$ open symbols) and (d) Phase angle as a function of frequency at different times during gelation: $t=1.21\h$ (\protect\circleG), $t=1.31\h$ (\protect\squareG), $t=t_g=1.46\h$ (\protect\triangleG) and $t=5.25\h$ (\protect\diamondG). Symbols and colors correspond to the points indicated in part (a) and (b). At the gel point the phase angle is invariant with frequency (\protect\triangleG), while before and after there is a more pronounced dependence of $\delta$ on $\omega$. The solid lines in part (c) and (d) corresponds to the model prediction obtained assuming a mechanical model composed of a single fractional element as described in section~\ref{sec:casein} (same color code applies); Part (e) and (f) show respectively the evolution in the gel strength and average relaxation exponent as a function of time. 
		\label{fig:figure4}}%
\end{figure*}

Casein gels are protein gels composed of sodium caseinate and deionized water. Their gelation can be induced by adding glucono-$\delta$-lactone (GDL), which spontaneously hydrolyzes in solution decreasing the pH. As soon as the solution isoelectric point is reached, agglomeration is induced and the caseinate proteins form an elastic sample-spanning network. The time at which gelation is initiated and the rate of mutation depend on the GDL content \cite{Braga2006}. Here, we present results for one specific composition with expected characteristic mutation time $\tau_{mu} \simeq 120 \s$ (see Appendix~\ref{app:materials} for more details). Before gelation begins, the solution is essentially a low-viscosity Newtonian fluid and can easily be transferred to the rheometer for measurements without any residual memory of loading history.

To quantify the casein gelation process with the ARES-G2 rheometer we monitor the acid-induced agglomeration of the colloidal proteins with a sequence of continuous OWCh signals. Immediately after preparation, the mixture is poured into a double gap Couette cell geometry, which has been prealigned and set in advance at $T=25\degC$. The rotor is then lowered to the working gap and the experimental procedure, already prepared in the rheometer software, is initiated.
The procedure consists of applying repeated OWCh signals over a duration of about $6$ hours. The force transducers are zeroed at the beginning of the experiment, while the strain is zeroed before each signal by imposing $\gamma = 0$ for a waiting time $t_w = 1\s$.
Given the estimated value of $\tau_{mu}$, we used chirp signals with the same length and frequency range used for the initial optimization process, i.e., $T = 14\s$, $\omega_1 = 0.3 \rads$, $\omega_2 = 30 \rads$, corresponding to $\TB \simeq 66$. The maximum strain was set to $\gamma_0 = 0.01$ and the tapering parameter $r=10\%$ for all signals. This allowed us to obtain the viscoelastic spectrum of the mutating casein gel over two decades of the angular frequency, while maintaining a mutation number much smaller than unity ($N_{mu} = T/\tau_{mu} = 0.1 $).
The resulting strain and stress signals from the rheometer are exported at the end of the experiment and processed separately.
The gelation dynamics are presented in Fig.~\ref{fig:figure4}(a)-(d), where
the first column (Fig.~\ref{fig:figure4}(a) and (b)) shows the evolution of the material behavior over time, while the second column (Fig.~\ref{fig:figure4}(c) and (d)) shows its dependence on the angular frequency for a few characteristic values of elapsed times.

In Fig.~\ref{fig:figure4}(a) the storage and loss moduli measured at $\omega = 10 \rads$ are plotted as a function of time, starting from the first input chirp. This figure is constructed by selecting the values of $G'$ and $G''$ obtained for $\omega = 10 \rads$ from all of the individual chirp signals and corresponds to a typical gelation test protocol. The data show that the mixture remains essentially Newtonian for about one hour after the experiment is started, with values of both moduli below the noise level of the instrument (gray data). During this time the pH of the mixture is slowly decreasing towards the isoelectric point. As soon as agglomeration is induced and the proteins begin to form an elastic network,  both the storage and loss moduli monotonically increase over time. There is a fast initial change over many decades in modulus that happens within the first half an hour from the beginning of the gelation process, followed by a subsequent slower evolution towards steady state.
The representation in Fig.~~]\ref{fig:figure4}(a) is equivalent to that obtained by conventional rheometric gelation experiments performed with a time sweep at a single constant $\omega$. Although for clarity only one frequency is shown here, using the OWCh protocol we actually have access to all frequencies between $\omega_{min} = 2 \pi/T = 0.45 \rads$ and $\omega_{max} = \omega_2 = 30 \rads$ at an interval $\Delta \omega = 2 \pi/T$, that is, more than $70$ frequencies for this specific experiment. Attempting to obtain the same results with classical time sweeps would dramatically decrease the number and range of frequencies accessible.

In Fig.~\ref{fig:figure4}(b) we present the evolution of the phase angle $\delta$ over time for three of the frequencies available. As in Fig.~\ref{fig:figure4}(a), there is an initial period where the torque signal is below the noise floor of the instrument. Then, after approximately one hour, a network begins to form which is always dominated by elasticity, since $\delta < \pi/4$ or equivalently $G' > G''$ at all times.
The GP can be clearly identified as the point in time (highlighted by a pink triangle at $t_{g} = 1.46 \;\mathrm{h}$) at which the curves at three different frequencies collapse, denoting the existence of a unique value of the phase angle $\delta$ independent of $\omega$. The other curves at different frequencies are not shown here for clarity but they all lie between the gray and the black bounding curves and also collapse at the GP.

In order to show more directly the frequency dependence of the viscoelastic moduli, in Fig.~\ref{fig:figure4}(c) we plot $G'$ and $G''$ as a function of the angular frequency at four different times during gelation, highlighted with matching symbols and colors in Figs.~\ref{fig:figure4}(a) and (b). In Fig.~\ref{fig:figure4}(d) the same data is presented in terms of the phase angle $\delta$. Figure~\ref{fig:figure4}(c) shows more directly that at all times, and for all the frequencies probed, this composition of the casein gel is dominated by elasticity with $G'>G''$.
In addition, the OWCh sequencing protocol allows us to detect a unique characteristic of this material which cannot be discerned by looking simply at the transient evolution of the moduli at one frequency: the measured storage and loss moduli are very close to a single power-law throughout the gelation process, not only at the exact gel point.
To quantify how closely the spectrum represents that of a critical gel, we look directly at the functional dependence of the phase angle on the angular frequency in Fig.~\ref{fig:figure4}(d). Thanks to the high time-frequency resolution of the OWCh signal we are able to distinguish a clear point in time (pink triangle at $t=t_g=1.46\,\mathrm{h}$) when the phase angle is a constant ($\delta_g = 0.34 \pm 0.01$) and which, within experimental error and sensitivity, uniquely identifies the GP. Before and after this critical point in time the functional form of $\delta(\omega)$ switches from a pre-gel form (monotonically decreasing with frequency), to a post-gel form (monotonically increasing with $\omega$); these characteristics have been previously reported for different mutating systems \cite{Curtis2014} and is considered a hallmark of the sol-to-gel transition. However, we can clearly see that, both around and after gelation, the differences in phase angles are less than $10\%$  over the entire range of frequencies tested (about two orders of magnitude). This particular protein gel system can therefore be considered as always being close to a critical gel.

The frequency resolution achieved using OWCh signals allows us not only to follow the evolution of the viscoelastic spectrum measured experimentally, but also gives us an opportunity to monitor how the material properties evolve over time once an appropriate constitutive model is selected. Since this particular protein gel is always very close to a critical gel, we model its evolution using a single mechanical fractional element with two time evolving quasi-properties $(\mathbb{V}(t),\alpha(t))$ \cite{Keshavarz2017}.
The relaxation modulus (or impulse response) of a single fractional element can be analytically derived \cite{Koeller1984} and is given by:
	\begin{equation}
		\label{eq:G_t}
		G(t) = \frac{\mathbb{V}}{\Gamma(1-\alpha)}t^{-\alpha}\, ,
	\end{equation}
where $\Gamma(s)$ is the Gamma function of the argument $s$. If we take the analytical Fourier transform of $G(t)$ we obtain an expression for the complex modulus (or transfer function) of the fractional element $G^\star(\omega) = G'(\omega) + i G''((\omega))$ with:
	\begin{align}
		\label{eq:gp}
		G'(\omega) &= \mathbb{V} \omega^\alpha \cos\left(  \frac{\pi}{2} \alpha \right) \, ,  \\
		\label{eq:gpp}
		G''(\omega) &= \mathbb{V} \omega^\alpha \sin\left( \frac{\pi}{2} \alpha \right) \, .
	\end{align}
Using these expressions we can then deduce that:
	\begin{align}
	\label{eq:alpha}
	\alpha &= \frac{2}{\pi} \tan^{-1} \left(\frac{G''}{G'} \right)\, ,  \\
	\label{eq:V}
	\mathbb{V} &= \frac{\sqrt{G'^2(\omega) + G''^2(\omega)}}{\omega^\alpha}\, .
	\end{align}
From Eqs.~(\ref{eq:alpha})~and~(\ref{eq:V}) we can easily calculate values of the model parameters for each frequency available and obtain a single unique average value at each time by averaging over all the frequencies tested $(\bar{\mathbb{V}}(t),\bar{\alpha} (t))$. We can then plot their evolution over the entire duration of the experiment as shown in Fig.~\ref{fig:figure4}(e) and (f) for the average gel strength $\bar{\mathbb{V}}(t)$ and the exponent $\bar{\alpha}(t)$ respectively. Using Eqs.~(\ref{eq:gp})~and~(\ref{eq:gpp}), together with the definition of $\delta$ and the values of $(\bar{\mathbb{V}}(t),\bar{\alpha}(t))$ estimated from the experimental data at the corresponding times, we can directly compare the model with the experimental data. The viscoelastic moduli and the phase angle thus calculated are plotted with solid lines (using the same color code) in Fig.~\ref{fig:figure4}(c) and (d) respectively.

This example illustrates the ability of the OWCh protocol to resolve the gelation process of a mutating sample with optimized time and frequency resolution, while guaranteeing a small mutation number and an imposed sample strain that always remains within the linear regime. The information attainable with this technique opens new possibilities to resolve macroscopic gelation dynamics of a wide range of systems  with a simple experimental protocol that can be performed on commercially available rheometer without the use of any additional hardware. 

%


\section{Summary}
\label{sec:summary}

Time-resolved mechanical spectroscopy is very important in the quest to relate microscopic dynamics to the bulk material behavior of soft materials. Such connections are fundamental to improving our understanding of a large number of soft materials currently employed in different industrial, biological and medical applications and will become even more essential in the development of the next generation of soft materials, with microstructural components designed to achieve specific macroscopic properties.

In this work we have addressed the issue of defining and optimizing the input signal for time-resolved mechanical spectroscopy. Based on the requirements specific to mechanical measurements, we have identified the potential of frequency modulated exponential chirp signals. These sequences are inherently affected by spectral leakage when working with small time-bandwidth constants. Inspired by the biosonar signals that bats and dolphins use for echolocation, we have tackled the issue of spectral leakage by designing a suitable modulation of the signal envelope using a Tukey window function, with adjustable time-width defined by a dimensionless parameter $r$. Using a stable reference PIB solution, we performed an optimization procedure which enabled us to identify an optimum range of values ($6 \le r \le 15 \%$) for which the error in the estimation of the viscoelastic moduli (defined with respect to the complex modulus measured by standard frequency sweeps) is decreased by almost two orders of magnitude. A comparison with detailed numerical simulations of the same material, using a fractional Maxwell model being sheared with the same windowed-chirp protocol, clearly shows that this optimal range is set by the extrinsic noise floor that is inevitably present in any experimental procedure. These experiments and computations highlight how using an optimally windowed chirp enables us to measure the linear viscoelastic spectrum with the same accuracy as a classical frequency sweep, while dramatically reducing the total measurement time by a factor of about $10^2$. This enables us to study the evolution in the viscoelastic response of time-evolving systems such as gels and thixotropic pastes.

To illustrate this, we used a sequence of OWCh signals to follow the gelation of an acid-induced protein solution with a characteristic mutation time of about $120 \s$. The length of the signals was chosen to guarantee a mutation number much smaller than unity. Analysis of the output stress and input strain time sequences allowed us to measure the viscoelastic moduli over two decades in frequency during the entire duration of the gelation process. The resulting information allows us to clearly discern the gel point and analyze the time-evolution of the material properties within the framework of a critical gel model that can be represented with a single fractional spring-pot element.

The OWCh framework constructed in this work is a powerful tool that will enable researchers across disciplines to study the mechanical macroscopic behavior of soft materials with both time and frequency resolution using current state-of-the-art rheometers, even with rapidly mutating systems. The signal characteristics can be adapted to different materials and instruments that are able to generate a strain- or stress- controlled arbitrary perturbation and can work in any frequency range accessible by the instrument. The ability to perform time-resolved mechanical spectroscopy at the macroscale, coupled with time-resolved microscopic probes such as velocimetry \cite{Gallot2013,SaintMichel2016}, light-scattering \cite{Cipelletti1999} and neutron-scattering \cite{Eberle2011}, opens novel possibilities for identifying the connection between the underlying microstructural dynamics and the bulk behavior that characterizes soft materials.

\appendix

\section{Materials and Methods}
\label{app:materials}
In order to develop and test the OWCh protocol we employed several different soft materials. The majority of initial experiments used to optimize the method were carried out on two reference fluids with a well defined viscoelastic spectrum: (i) a semi-dilute polymer solution of $8.5 \,\mathrm{wt\%}$ poly(isobutylene) in hexadecane (both supplied by Sigma Aldrich), referred to as a PIB solution in the main text, and (ii) a surfactant-counterion solution of
cetylpyridinium chloride (CPyCl), sodium salicilate (NaSal) and sodium cloride (NaCl) in DI water with CPyCl:NaSal:NaCl concentrations of 100:60:$33  \, \mathrm{mM}$, referred to as a micellar solution (CPyCl and NaSal supplied by Alfa Aesar, reagent grade NaCl was purchased from Sigma Aldrich). The storage and loss moduli of the PIB solution measured with a standard frequency sweep are shown in Fig.~\ref{fig:figure1}c by the black solid and dashed lines respectively. The storage modulus $G'$ is lower than the loss modulus $G''$ for frequencies below the cross over frequency $\omega_c = 30 \rads$, meaning that in the limit of small deformation rates the material flows similarly to a viscous fluid. However, for frequencies with $\omega > \omega_c$ the material becomes increasingly elastic and $G'$ becomes greater than $G''$. This type of linear viscoelastic response is prototypical of many complex fluids as well as a mutating system in its pre-gel state, i.e., before gelation has occurred.

The PIB solution is designed to emulate the characteristics of a NIST Standard Reference Material (SRM1490) even though this sample is no longer readily available or supported \cite{Schultheisz2000}. It is very stable over time and does not evaporate even after several days of experiments, making it an ideal candidate to test different signal sequences without incurring errors related to slow changes in composition. The aqueous micellar solution, on the other hand, is more volatile and therefore each sample was used for no longer than $40 \minutes$ each time, as determined by initial exploratory tests. The viscoelastic spectrum for this material is discussed in more detail in Appendix~\ref{app:micellar_solution}.
%

A third material was also tested using the optimized signal to provide an example of time-resolved mechanical spectroscopy on a mutating system, in this case an acid-induced protein gel \cite{Totosaus2002}. The gel was prepared by dissolving caseinate powder (Firmenich) at $4\%$ wt in deionized water. Homogeneous gelation was induced by dissolving $1\%$ wt glucono-$\delta$-lactone (GDL, Firmenich) into the protein solution. Gelation does not begin immediately after dissolution, leaving time to transfer this initially Newtonian mixture to the rheometer in its liquid state with no memory of its loading history.

All the measurements were carried out on a strain controlled ARES-G2 rheometer (TA Instruments). The PIB solution was tested with a  $50 \mm$ diameter, $1^\circ$ stainless steel cone, the bottom plate being directly connected to a Peltier stage, allowing for precise control of the sample temperature that was maintained at $25 \degC$ during all experiments. The micellar solution was tested using a $40 \mm$, $2^\circ$ stainless steel cone and the same lower Peltier plate. This cone is equipped with a solvent trap that was filled with water during the experiments and the sample enclosed by the solvent trap cover to minimize evaporation. The casein gel was tested in a double gap concentric cylinder geometry with a $32 \mm$ diameter recessed end bob and $34 \mm$ diameter cup, both constructed from anodized aluminum. Double gap geometries have the largest surface area in contact with the fluid and therefore result in a larger torque signal throughout the gelation process.



All the experimental data acquired with the ARES-G2 rheometer for different chirp sequences were exported and post-processed separately in \textsc{Matlab}. The signals, measured in time with an acquisition frequency $f_s = 500 \Hz$ for all the experiments presented in this work, were first checked for any DC bias in the baseline average value due to either the initial motor position or an inaccurate zeroing of the torque transducer, and then shifted to correct for such bias whenever necessary. After several experimental and numerical trials, this procedure was found to be essential in order to obtain the most accurate values of the viscoelastic moduli. We later found that a similar suggestion had been reported by John W. Tukey and coworkers \cite{Blackman1959book} as well. Before being processed via the in-built FFT functions, each signal was also cut to eliminate the initial waiting time $t_w$ and then zero-padded after the last data point to extend the signal to the closest power of two to improve the performance of FFT algorithms.
Time-dependent data generated with highly resolved numerical simulations were also down-sampled to the same acquisition frequency and processed using the exact same functions written for experimental data, the only difference being the unnecessary step of checking for possible biases.

Based on the files written for the analysis of the data presented in this work, we have developed a \textsc{Matlab} GUI to facilitate the application of the OWCh techique. The GUI has two main sections: one to generate the signal in the time domain that is then used as an input in the rheometer, and one to perform the steps for post-processing the measured data and extract the linear viscoelastic response $G^\star (\omega) = G'(\omega)+iG''(\omega)$.
The GUI is free and available upon request to the following email address \textcolor{red}{owch@mit.edu}.

\section{Micellar solution}
\label{app:micellar_solution}

\begin{figure}
	\centering
	\includegraphics[width=0.8\columnwidth,trim={0.6cm 0 0 0}]{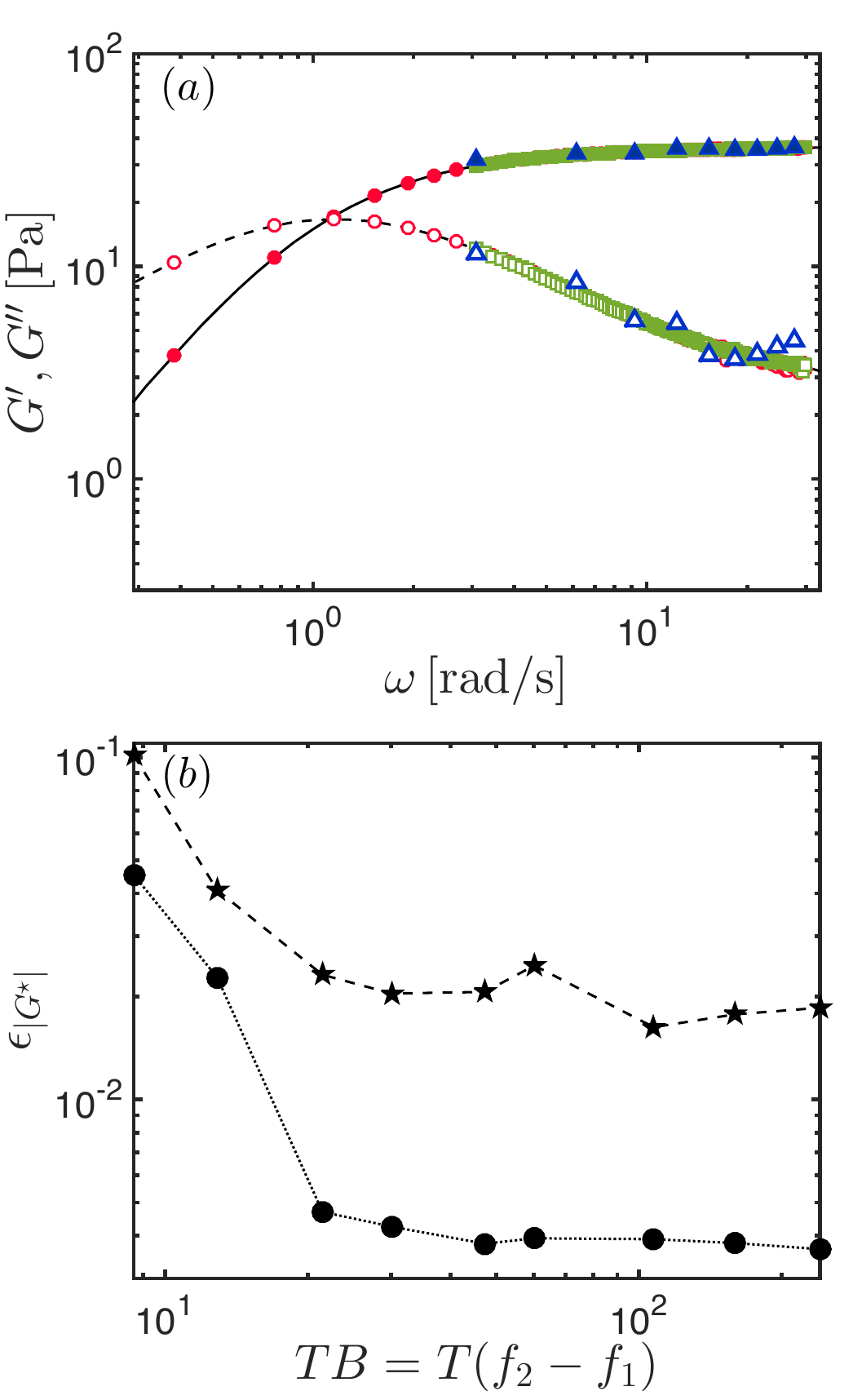}
	\caption{(a) Viscoelastic moduli of the CPyCl micellar solution. Reference values ($G'$ \protect\Solidline, $G''$ \protect\dashedline) were measured with a sequence of sine steps at different frequencies . Comparison between results measured using optimally windowed chirps ($r=10\%$) with two different time-bandwidth products obtained by changing only the length of the signal $T$: $\TB=8.6$ ($G'$ \protect\triangleH, $G''$ \protect\triangleHw) and $\TB=107$ ($G'$ \protect\squareH, $G''$ \protect\squareHw). The frequency range is kept constant with $\omega_1 = 3\rads$ and $\omega_2 = 30\rads$ to guarantee a consistent comparison between chirps of different durations. Results obtained using a separate optimally windowed chirp ($r=10\%$) with the frequency range adjusted to the length of the signal ($T=14\s$, $\omega_1 = 0.3\rads$, $\omega_2 = 30\rads$, $\TB = 66$) are also reported with circles ($G'$ \protect\circleH, $G''$ \protect\circleHw). The nominal strain for all signals (sine steps and chirps) is $\gamma_0 = 6\%$. ; (b) Trends of the error for the magnitude of the complex modulus as a function of varying time-bandwidth constant $\TB$ for an optimally windowed chirp with $r=10\%$. Results for both the PIB solution (\protect\circleI) and the micellar solution (\protect\StarI) are shown.
		\label{fig:figure_SI_1}}%
\end{figure}

As mentioned in section~\ref{subsec:optimization}, the experimental procedure used to determine the optimal value of the tapering parameter was employed on both the PIB solution and also on the aqueous micellar solution. The viscoelastic moduli as measured with a conventional discrete frequency sweep are shown in Fig.~\ref{fig:figure_SI_1}(a) in black. As explained in section~\ref{subsec:PIBconst_eq}, the micellar solution is almost perfectly described by a simple single mode Maxwell model (with $G=38 \Pa$ and $\tau=\eta/G = 1.19 \s$) over the range of frequencies of interest. Figure~\ref{fig:figure_SI_1}(a) also shows the values of $G'$ and $G''$ obtained using two different windowed chirps, both of them with $r=10\%$ but with two different values of the time-bandwidth constant.
In order to systematically compare the results, the moduli are shown only within the frequency range that is resolvable with the smallest $\TB$ tested, i.e., $\TB = 8.6$. The chirps were designed to span one decade in frequency with $\omega_1 = 3 \rads$, $\omega_2 = 30 \rads$ for any time-bandwidth constant, this way the shortest signal can be imposed as fast as $T = 2 \pi /\omega_1 \simeq 2 \s$. The other signals were then adjusted increasing their length while maintaining the same frequency bandwidth, thus changing the value of $\TB$. We can clearly see from Fig.~\ref{fig:figure_SI_1}(b) that using a larger value of the time-bandwidth constant is beneficial in terms of reducing spectral leakage even when using an optimal amount of tapering. Separate tests (not shown here) demonstrate that this effect is even more pronounced if the window function is not optimized, in which case one would need to have $\TB \gg 100$ to reach a comparable level of accuracy. In fact, from Fig.~\ref{fig:figure_SI_1}(b) we can see that the error in the magnitude of the complex modulus (defined analogously to the error for $G'$ and $G''$ given by Eqs.~(\ref{eq:error_gp}) and~(\ref{eq:error_gpp})) reaches a minimum plateau as soon as $\TB >10$ when using OWCh sequences, allowing us to extend the use of windowed chirps to very short signals (with $T\simeq 2 \s$) that would otherwise give unreliable estimations of the moduli. We note that for values of the time-bandwidth constant in the range $[10,100]$ it is almost impossible to obtain an estimate of the relaxation spectrum over multiple frequencies using a sequence of individual sine steps while maintaining a sufficiently low mutation number.

As anticipated, both the PIB solution and the micellar solution show the same trends in reduction of error with increasing $\TB$ as highlighted in Fig.~\ref{fig:figure_SI_1}(b).
However, since the complex modulus of the micellar fluid is much smaller in magnitude compared to that of the PIB solution, the smaller signal-to-noise means that the minimum error attainable using an optimally windowed chirp on the micellar solution is not as low as that for the PIB solution. This is also the case for classical frequency sweeps since the precision of both techniques is lower for smaller SNR.

\section{Error Analysis}
\label{app:error_analysis}

In order to achieve a more quantitative analysis of the RMS error as a function of the tapering parameter $r$ presented in section~\ref{subsec:optimization}, we provide here estimates for the error based on simple scaling arguments.
To show the effect of the window shape, we also consider an extended family of Tukey windows by assuming that the trigonometric term defining the amplitude modulation in Eq.~(\ref{eq:window}) can be elevated to any integer power $n$. Although not commonly used, in the limit of $r=100 \%$ such windows correspond to the well characterized family of sine/cosine window functions \cite{Harris1978}. In the following, we will refer to them as Tukey windows with degree $n$; for example, the original Tukey window introduced in section~\ref{subsec:windowed_chirp} is of degree $2$.

\begin{figure}[h]
	\centering
	\includegraphics[trim = 0mm 40mm 0mm 30mm, clip,width=\columnwidth]{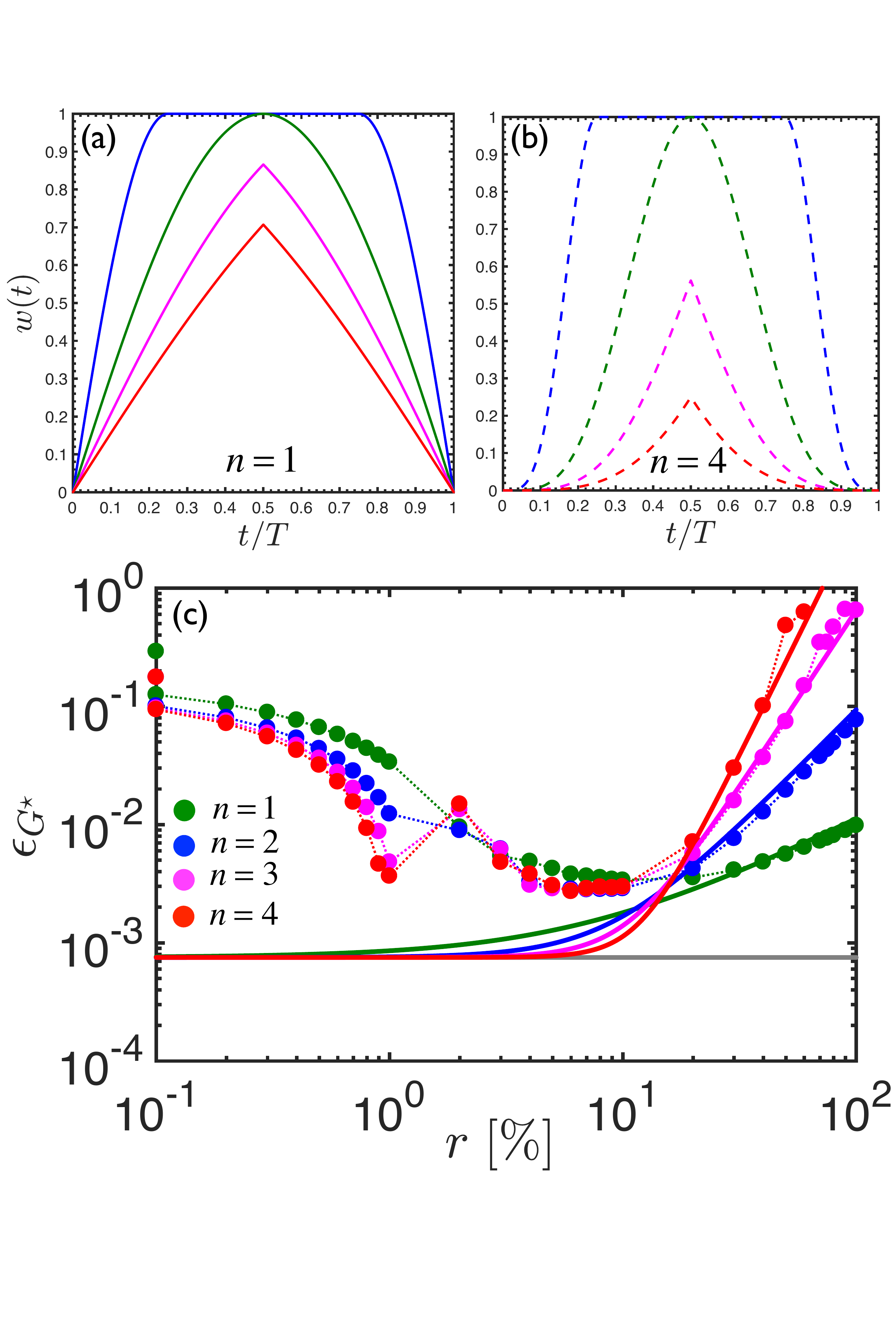}
	\caption{The window function $w(t)$ is plotted for Tukey windows with (a) $n=1$ and (b) $n=4$. The different blue, green, magenta and red colors correspond to $r=50\%$ (\protect\solidlineb), $r=100\%$ (\protect\solidlineg), $r=150\%$  (\protect\solidlinem) and $r=200\%$ (\protect\solidliner) respectively; (c) The computed errors for the complex modulus, from numerical simulations, are plotted versus the tapering parameter $r$ for different Tukey windows: $n=1$ (\protect\circleg), $n=2$ (\protect\circleb ), $n=3$ (\protect\circlem) and $n=4$ (\protect\circler). The corresponding solid lines show the error predictions from Eq.~(\ref{AppCeqn4}) with $c_1=0.3$ and $c_2=0.126$. The gray line represents the case of $n=0$.
		\label{fig:figure_SI_2}}%
\end{figure}

As discussed previously, in an ideal system with zero noise (in the output signal) the error in the measured modulus decreases by tapering the input signal due to a lower level of spectral leakage. However, real rheometric systems have a certain noise level in their stress signal which leads to a corresponding noise level present in the Fourier domain, $\tilde{\sigma}_{\text{noise}}(\omega)$.
As the tapering of the strain signal increases, the amplitude of the raw strain and the resulting stress signal decreases and ultimately becomes comparable to the noise level. It is known  that the corresponding error from noise in the measurements scales as \cite{Pintelon2012}:
\begin{equation}\label{AppCeqn1}
\epsilon_{\text{noise}}(\omega) \simeq \frac{1}{\mathrm{SNR} (\omega)} = \left|\frac{\tilde{\sigma}_{\text{noise}}(\omega)}{G^\star(\omega)\tilde{\gamma}(\omega)}\right|.
\end{equation}
While low levels of tapering (small values of $r$) reduce the spectral leakage (and its corresponding error) with a negligible effect on the SNR, for higher tapering levels (large values of $r$) the reduction in the power of the input signal leads to a significant increase in the noise/signal ratio (or equivalently a substantial decrease in the SNR, as evident from Eq.~(\ref{AppCeqn1})).
As shown in Fig.~\ref{fig:figure_SI_2}(a) and (b), the amplitude of the strain signal decreases dramatically as the tapering ratio $r$ approaches unity. This is true for all different degrees Tukey windows
although it is more accentuated for larger values of $n$.
One can clearly observe this fact by comparing the corresponding Tukey windows, at similar $r$ values, in Fig.~\ref{fig:figure_SI_2}(a) and (b) for $n=1$ and $n=4$ respectively.

To characterize this adverse effect of tapering at large values of $r$, we use the stationary phase method (\cite{Cheng2007,Bender2013}) and find the following estimate for the Fourier transform of a windowed chirp signal:
\begin{equation}\label{AppCeqn2}
\tilde{\gamma}(\omega)\simeq\gamma_0\sqrt{2\pi}\frac{w(t_k,r)}{\sqrt{\left(\omega/T\right)\log(\omega_2/\omega_1)}}
\end{equation}
where $t_k=T\log(\omega/\omega_1)/\log(\omega_2/\omega_1)$. Combining Eqs.~(\ref{AppCeqn1}) and~(\ref{AppCeqn2}), it is evident that the error is maximum when the window function and the complex modulus are both minimum. Thus, we can analyze the error close to the lowest frequency $\omega=\omega_1$ for which $t_k=0$. Again, by using the method of stationary phase and expanding the integrand to the first non-zero order, we find the following asymptotes for the Fourier transform of the strain signal:
\begin{equation}\label{AppCeqn3}
\tilde{\gamma}(\omega_1)\simeq\gamma_0 a(n)T\left(\frac{\pi}{r}\right)^n\left(T\omega_1\log\frac{\omega_2}{\omega_1}\right)^{-(n+1)/2}
\end{equation}
where $n$ corresponds to the degree of the Tukey window. The pre-factor $a(n)$ is equal to $\sqrt{2\pi}/2$, $1$, $\sqrt{2\pi}/2$, $2$, and $3\sqrt{2\pi}/2$ for integer values of $n$ from $0$ to $4$ respectively. This scaling for $\tilde{\gamma}(\omega_1)$, along with the scaling suggested in Eq.~\ref{AppCeqn1}, suggests that the maximum error due to the noise in the signal scales as $\tilde{\sigma}_{\text{noise}}(\omega_1)/G^\star(\omega_1)\tilde{\gamma}(\omega_1)$ which grows with the tapering ratio in a power-law manner $\sim r^n$. However, the average error is calculated for a combination of data points which are both in and out of the tapered zone. Consequently, we propose that the average error is the following linear superposition of error for the internal points (which, similar to the $n=0$ case, are not tapered by the window) and the points at the start and end of the signal (that are tapered by a Tukey window with degree $n$):
\begin{equation}\label{AppCeqn4}
\epsilon_{G^\star}=c_1\left(\frac{\tilde{\sigma}_{\text{noise}}(\omega_1)}{NG^\star(\omega_1)\gamma_0}\right)\left[\frac{2\sqrt{A}}{\sqrt{2\pi}}+\frac{1}{a(n)}\left(\frac{r}{c_2\pi}\right)^nA^{(n+1)/2}\right]
\end{equation}
where $A=T\omega_1\log({\omega_2}/{\omega_1})$, $N$ is the number of points in the frequency domain over which the average error is calculated, and $c_1$ and $c_2$ are two fitting constants of order unity. It is important to emphasize that the proposed formula only captures the error due to the noise level in the stress signal and does not include the contributions due to the spectral leakage. However, at large values of $r$ the error due to spectral leakage is minimal and the average error is dominated by the transducer noise in the stress signal (as supported by our numerical simulations in section~\ref{sec:numerical_simulations}). In this limit, as the proposed formula suggests, the average error increases with the tapering parameter in a power-law manner $\epsilon_{G^\star}\sim r^n$.\\
In order to check the proposed prediction, we performed additional numerical simulations following the procedure described in section~\ref{sec:numerical_simulations}. We changed the degree of the applied Tukey windows and performed tests with window families at four different values of $n=1,2,3,4$. Figure~\ref{fig:figure_SI_2}(c) shows the computed values of the RMS error in the complex modulus $\epsilon_{G^\star}$ as a function of the tapering parameter $r$ for different degrees of Tukey windows. It is evident that at large values of the tapering ratio ($r\geq0.1$), the error scales as $r^n$.\\
The proposed formula for the average error (Eq.~(\ref{AppCeqn4})) is fitted to the data and for all tested families of Tukey windows with different values of $n$, the predicted error (Eq.~(\ref{AppCeqn4}) with $c_1=0.3$ and $c_2=0.126$) matches well with the computed RMS errors from the numerical simulations at large values of $r$ ($r\ge 10\%$).

This analysis further supports the conclusions drawn from section~\ref{sec:numerical_simulations} that, at large values of $r$, the background noise in the measured signal can induce significant errors which grow with the tapering parameter $r$. It also shows that such errors increase in a polynomial manner with the degree of the window function $n$. By contrast, at small values of the tapering parameter ($r\leq 10\%$) the error is dominated by spectral leakage and decreases with increases in the tapering ratio.
At intermediate values of $r \in [6,15] \%$ these two contributions to the noise both play a role and constrain the maximum reduction in noise that can be achieved via the OWCh protocol.
The comparison of different Tukey windows shown in Fig.~\ref{fig:figure_SI_2}(c) also highlights that using a degree $n=2$ is advantageous. In fact, while the initial decrease in the error is comparable to that of higher degrees (with $n=1$ having the highest values of $\varepsilon_{G^\star}$ for the same $r$), the subsequent divergence is the least pronounced and thus provides the widest range of $r$ for maximum noise reduction.   \\


\begin{thebibliography}{80}%
		\makeatletter
		\providecommand \@ifxundefined [1]{%
			\@ifx{#1\undefined}
		}%
		\providecommand \@ifnum [1]{%
			\ifnum #1\expandafter \@firstoftwo
			\else \expandafter \@secondoftwo
			\fi
		}%
		\providecommand \@ifx [1]{%
			\ifx #1\expandafter \@firstoftwo
			\else \expandafter \@secondoftwo
			\fi
		}%
		\providecommand \natexlab [1]{#1}%
		\providecommand \enquote  [1]{``#1''}%
		\providecommand \bibnamefont  [1]{#1}%
		\providecommand \bibfnamefont [1]{#1}%
		\providecommand \citenamefont [1]{#1}%
		\providecommand \href@noop [0]{\@secondoftwo}%
		\providecommand \href [0]{\begingroup \@sanitize@url \@href}%
		\providecommand \@href[1]{\@@startlink{#1}\@@href}%
		\providecommand \@@href[1]{\endgroup#1\@@endlink}%
		\providecommand \@sanitize@url [0]{\catcode `\\12\catcode `\$12\catcode
			`\&12\catcode `\#12\catcode `\^12\catcode `\_12\catcode `\%12\relax}%
		\providecommand \@@startlink[1]{}%
		\providecommand \@@endlink[0]{}%
		\providecommand \url  [0]{\begingroup\@sanitize@url \@url }%
		\providecommand \@url [1]{\endgroup\@href {#1}{\urlprefix }}%
		\providecommand \urlprefix  [0]{URL }%
		\providecommand \Eprint [0]{\href }%
		\providecommand \doibase [0]{http://dx.doi.org/}%
		\providecommand \selectlanguage [0]{\@gobble}%
		\providecommand \bibinfo  [0]{\@secondoftwo}%
		\providecommand \bibfield  [0]{\@secondoftwo}%
		\providecommand \translation [1]{[#1]}%
		\providecommand \BibitemOpen [0]{}%
		\providecommand \bibitemStop [0]{}%
		\providecommand \bibitemNoStop [0]{.\EOS\space}%
		\providecommand \EOS [0]{\spacefactor3000\relax}%
		\providecommand \BibitemShut  [1]{\csname bibitem#1\endcsname}%
		\let\auto@bib@innerbib\@empty
		\bibitem [{\citenamefont {Mezzenga}\ \emph {et~al.}(2005)\citenamefont
			{Mezzenga}, \citenamefont {Schurtenberger}, \citenamefont {Burbidge},\ and\
			\citenamefont {Michel}}]{Mezzenga2005}%
		\BibitemOpen
		\bibfield  {author} {\bibinfo {author} {\bibfnamefont {R.}~\bibnamefont
				{Mezzenga}}, \bibinfo {author} {\bibfnamefont {P.}~\bibnamefont
				{Schurtenberger}}, \bibinfo {author} {\bibfnamefont {A.}~\bibnamefont
				{Burbidge}}, \ and\ \bibinfo {author} {\bibfnamefont {M.}~\bibnamefont
				{Michel}},\ }\href {\doibase 10.1038/nmat1496} {\bibfield  {journal}
			{\bibinfo  {journal} {Nature Materials}\ }\textbf {\bibinfo {volume} {4}},\
			\bibinfo {pages} {729} (\bibinfo {year} {2005})}\BibitemShut {NoStop}%
		\bibitem [{\citenamefont {Chen}\ \emph {et~al.}(2010)\citenamefont {Chen},
			\citenamefont {Wen}, \citenamefont {Janmey}, \citenamefont {Crocker},\ and\
			\citenamefont {Yodh}}]{Chen2010}%
		\BibitemOpen
		\bibfield  {author} {\bibinfo {author} {\bibfnamefont {D.~T.~N.}\
				\bibnamefont {Chen}}, \bibinfo {author} {\bibfnamefont {Q.}~\bibnamefont
				{Wen}}, \bibinfo {author} {\bibfnamefont {P.~A.}\ \bibnamefont {Janmey}},
			\bibinfo {author} {\bibfnamefont {J.~C.}\ \bibnamefont {Crocker}}, \ and\
			\bibinfo {author} {\bibfnamefont {A.~G.}\ \bibnamefont {Yodh}},\ }\href
		{\doibase 10.1146/annurev-conmatphys-070909-104120} {\bibfield  {journal}
			{\bibinfo  {journal} {Annual Review of Condensed Matter Physics}\ }\textbf
			{\bibinfo {volume} {1}},\ \bibinfo {pages} {301} (\bibinfo {year}
			{2010})}\BibitemShut {NoStop}%
		\bibitem [{\citenamefont {Gibaud}\ \emph {et~al.}(2012)\citenamefont {Gibaud},
			\citenamefont {Mahmoudi}, \citenamefont {Oberdisse}, \citenamefont {Lindner},
			\citenamefont {Pedersen}, \citenamefont {Oliveira}, \citenamefont
			{Stradner},\ and\ \citenamefont {Schurtenberger}}]{Gibaud2012}%
		\BibitemOpen
		\bibfield  {author} {\bibinfo {author} {\bibfnamefont {T.}~\bibnamefont
				{Gibaud}}, \bibinfo {author} {\bibfnamefont {N.}~\bibnamefont {Mahmoudi}},
			\bibinfo {author} {\bibfnamefont {J.}~\bibnamefont {Oberdisse}}, \bibinfo
			{author} {\bibfnamefont {P.}~\bibnamefont {Lindner}}, \bibinfo {author}
			{\bibfnamefont {J.~S.}\ \bibnamefont {Pedersen}}, \bibinfo {author}
			{\bibfnamefont {C.~L.~P.}\ \bibnamefont {Oliveira}}, \bibinfo {author}
			{\bibfnamefont {A.}~\bibnamefont {Stradner}}, \ and\ \bibinfo {author}
			{\bibfnamefont {P.}~\bibnamefont {Schurtenberger}},\ }\href {\doibase
			10.1039/c2fd20048a} {\bibfield  {journal} {\bibinfo  {journal} {Faraday
					Discussions}\ }\textbf {\bibinfo {volume} {158}},\ \bibinfo {pages} {267}
			(\bibinfo {year} {2012})}\BibitemShut {NoStop}%
		\bibitem [{\citenamefont {Chan}\ \emph {et~al.}(2016)\citenamefont {Chan},
			\citenamefont {Low}, \citenamefont {Heng}, \citenamefont {Chan},
			\citenamefont {Owh},\ and\ \citenamefont {Loh}}]{Chan2016}%
		\BibitemOpen
		\bibfield  {author} {\bibinfo {author} {\bibfnamefont {B.~Q.~Y.}\
				\bibnamefont {Chan}}, \bibinfo {author} {\bibfnamefont {Z.~W.~K.}\
				\bibnamefont {Low}}, \bibinfo {author} {\bibfnamefont {S.~J.~W.}\
				\bibnamefont {Heng}}, \bibinfo {author} {\bibfnamefont {S.~Y.}\ \bibnamefont
				{Chan}}, \bibinfo {author} {\bibfnamefont {C.}~\bibnamefont {Owh}}, \ and\
			\bibinfo {author} {\bibfnamefont {X.~J.}\ \bibnamefont {Loh}},\ }\href
		{\doibase 10.1021/acsami.6b01295} {\bibfield  {journal} {\bibinfo  {journal}
				{ACS Applied Materials {\&} Interfaces}\ }\textbf {\bibinfo {volume} {8}},\
			\bibinfo {pages} {10070} (\bibinfo {year} {2016})}\BibitemShut {NoStop}%
		\bibitem [{\citenamefont {Gong}\ and\ \citenamefont {Osada}(2010)}]{Gong2010}%
		\BibitemOpen
		\bibfield  {author} {\bibinfo {author} {\bibfnamefont {J.~P.}\ \bibnamefont
				{Gong}}\ and\ \bibinfo {author} {\bibfnamefont {Y.}~\bibnamefont {Osada}},\
		}in\ \href {\doibase 10.1007/12_2010_91} {\emph {\bibinfo {booktitle}
				{Advances in Polymer Science}}},\ Vol.\ \bibinfo {volume} {236},\ \bibinfo
		{editor} {edited by\ \bibinfo {editor} {\bibfnamefont {M.}~\bibnamefont
				{Cloitre}}}\ (\bibinfo  {publisher} {Springer Berlin Heidelberg},\ \bibinfo
		{year} {2010})\ pp.\ \bibinfo {pages} {203--246}\BibitemShut {NoStop}%
		\bibitem [{\citenamefont {Flory}(1953)}]{Flory1953}%
		\BibitemOpen
		\bibfield  {author} {\bibinfo {author} {\bibfnamefont {P.~J.}\ \bibnamefont
				{Flory}},\ }\href@noop {} {\emph {\bibinfo {title} {{Principles of polymer
						chemistry}}}}\ (\bibinfo  {publisher} {Cornell University Press},\ \bibinfo
		{year} {1953})\BibitemShut {NoStop}%
		\bibitem [{\citenamefont {Kavanagh}\ and\ \citenamefont
			{Ross-Murphy}(1998)}]{Kavanagh1998}%
		\BibitemOpen
		\bibfield  {author} {\bibinfo {author} {\bibfnamefont {G.~M.}\ \bibnamefont
				{Kavanagh}}\ and\ \bibinfo {author} {\bibfnamefont {S.~B.}\ \bibnamefont
				{Ross-Murphy}},\ }\href {\doibase
			https://doi.org/10.1016/S0079-6700(97)00047-6} {\bibfield  {journal}
			{\bibinfo  {journal} {Progress in Polymer Science}\ }\textbf {\bibinfo
				{volume} {23}},\ \bibinfo {pages} {533} (\bibinfo {year} {1998})}\BibitemShut
		{NoStop}%
		\bibitem [{\citenamefont {Osada}\ \emph {et~al.}(2004)\citenamefont {Osada},
			\citenamefont {Gong},\ and\ \citenamefont {Tanaka}}]{Osada2004}%
		\BibitemOpen
		\bibfield  {author} {\bibinfo {author} {\bibfnamefont {Y.}~\bibnamefont
				{Osada}}, \bibinfo {author} {\bibfnamefont {J.~P.}\ \bibnamefont {Gong}}, \
			and\ \bibinfo {author} {\bibfnamefont {Y.}~\bibnamefont {Tanaka}},\ }\href
		{\doibase 10.1081/MC-120027935} {\bibfield  {journal} {\bibinfo  {journal}
				{Journal of Macromolecular Science - Polymer Reviews}\ }\textbf {\bibinfo
				{volume} {44}},\ \bibinfo {pages} {87} (\bibinfo {year} {2004})}\BibitemShut
		{NoStop}%
		\bibitem [{\citenamefont {Bremer}\ \emph {et~al.}(1990)\citenamefont {Bremer},
			\citenamefont {Bijsterbosch}, \citenamefont {Schrijvers}, \citenamefont {van
				Vliet},\ and\ \citenamefont {Walstra}}]{Bremer1990}%
		\BibitemOpen
		\bibfield  {author} {\bibinfo {author} {\bibfnamefont {L.~G.}\ \bibnamefont
				{Bremer}}, \bibinfo {author} {\bibfnamefont {B.~H.}\ \bibnamefont
				{Bijsterbosch}}, \bibinfo {author} {\bibfnamefont {R.}~\bibnamefont
				{Schrijvers}}, \bibinfo {author} {\bibfnamefont {T.}~\bibnamefont {van
					Vliet}}, \ and\ \bibinfo {author} {\bibfnamefont {P.}~\bibnamefont
				{Walstra}},\ }\href {\doibase 10.1016/0166-6622(90)80139-U} {\bibfield
			{journal} {\bibinfo  {journal} {Colloids and Surfaces}\ }\textbf {\bibinfo
				{volume} {51}},\ \bibinfo {pages} {159} (\bibinfo {year} {1990})}\BibitemShut
		{NoStop}%
		\bibitem [{\citenamefont {Totosaus}\ \emph {et~al.}(2002)\citenamefont
			{Totosaus}, \citenamefont {Montejano}, \citenamefont {Salazar},\ and\
			\citenamefont {Guerrero}}]{Totosaus2002}%
		\BibitemOpen
		\bibfield  {author} {\bibinfo {author} {\bibfnamefont {A.}~\bibnamefont
				{Totosaus}}, \bibinfo {author} {\bibfnamefont {J.~G.}\ \bibnamefont
				{Montejano}}, \bibinfo {author} {\bibfnamefont {J.~A.}\ \bibnamefont
				{Salazar}}, \ and\ \bibinfo {author} {\bibfnamefont {I.}~\bibnamefont
				{Guerrero}},\ }\href {\doibase 10.1046/j.1365-2621.2002.00623.x} {\bibfield
			{journal} {\bibinfo  {journal} {International Journal of Food Science and
					Technology}\ }\textbf {\bibinfo {volume} {37}},\ \bibinfo {pages} {589}
			(\bibinfo {year} {2002})}\BibitemShut {NoStop}%
		\bibitem [{\citenamefont {Leocmach}\ \emph {et~al.}(2014)\citenamefont
			{Leocmach}, \citenamefont {Perge}, \citenamefont {Divoux},\ and\
			\citenamefont {Manneville}}]{Leocmach2014}%
		\BibitemOpen
		\bibfield  {author} {\bibinfo {author} {\bibfnamefont {M.}~\bibnamefont
				{Leocmach}}, \bibinfo {author} {\bibfnamefont {C.}~\bibnamefont {Perge}},
			\bibinfo {author} {\bibfnamefont {T.}~\bibnamefont {Divoux}}, \ and\ \bibinfo
			{author} {\bibfnamefont {S.}~\bibnamefont {Manneville}},\ }\href {\doibase
			10.1103/PhysRevLett.113.038303} {\bibfield  {journal} {\bibinfo  {journal}
				{Physical Review Letters}\ }\textbf {\bibinfo {volume} {113}},\ \bibinfo
			{pages} {038303, 1} (\bibinfo {year} {2014})}\BibitemShut {NoStop}%
		\bibitem [{\citenamefont {Keshavarz}\ \emph {et~al.}(2017)\citenamefont
			{Keshavarz}, \citenamefont {Divoux}, \citenamefont {Manneville},\ and\
			\citenamefont {McKinley}}]{Keshavarz2017}%
		\BibitemOpen
		\bibfield  {author} {\bibinfo {author} {\bibfnamefont {B.}~\bibnamefont
				{Keshavarz}}, \bibinfo {author} {\bibfnamefont {T.}~\bibnamefont {Divoux}},
			\bibinfo {author} {\bibfnamefont {S.}~\bibnamefont {Manneville}}, \ and\
			\bibinfo {author} {\bibfnamefont {G.~H.}\ \bibnamefont {McKinley}},\ }\href
		{\doibase 10.1021/acsmacrolett.7b00213} {\bibfield  {journal} {\bibinfo
				{journal} {ACS Macro Letters}\ }\textbf {\bibinfo {volume} {6}},\ \bibinfo
			{pages} {663} (\bibinfo {year} {2017})}\BibitemShut {NoStop}%
		\bibitem [{\citenamefont {Zaccarelli}(2007)}]{Zaccarelli2007}%
		\BibitemOpen
		\bibfield  {author} {\bibinfo {author} {\bibfnamefont {E.}~\bibnamefont
				{Zaccarelli}},\ }\href {\doibase 10.1088/0953-8984/19/32/323101} {\bibfield
			{journal} {\bibinfo  {journal} {Journal of Physics Condensed Matter}\
			}\textbf {\bibinfo {volume} {19}},\ \bibinfo {pages} {323101, 1} (\bibinfo
			{year} {2007})}\BibitemShut {NoStop}%
		\bibitem [{\citenamefont {Trappe}\ and\ \citenamefont
			{Weitz}(2000)}]{Trappe2000}%
		\BibitemOpen
		\bibfield  {author} {\bibinfo {author} {\bibfnamefont {V.}~\bibnamefont
				{Trappe}}\ and\ \bibinfo {author} {\bibfnamefont {D.~A.}\ \bibnamefont
				{Weitz}},\ }\href {\doibase https://doi.org/10.1103/PhysRevLett.85.449}
		{\bibfield  {journal} {\bibinfo  {journal} {Physical Review Letters}\
			}\textbf {\bibinfo {volume} {85}},\ \bibinfo {pages} {449} (\bibinfo {year}
			{2000})}\BibitemShut {NoStop}%
		\bibitem [{\citenamefont {Lu}\ and\ \citenamefont {Weitz}(2013)}]{Lu2013}%
		\BibitemOpen
		\bibfield  {author} {\bibinfo {author} {\bibfnamefont {P.~J.}\ \bibnamefont
				{Lu}}\ and\ \bibinfo {author} {\bibfnamefont {D.~A.}\ \bibnamefont {Weitz}},\
		}\href {\doibase 10.1146/annurev-conmatphys-030212-184213} {\bibfield
			{journal} {\bibinfo  {journal} {Annual Review of Condensed Matter Physics}\
			}\textbf {\bibinfo {volume} {4}},\ \bibinfo {pages} {217} (\bibinfo {year}
			{2013})}\BibitemShut {NoStop}%
		\bibitem [{\citenamefont {Liu}\ and\ \citenamefont {Nagel}(1998)}]{Liu1998}%
		\BibitemOpen
		\bibfield  {author} {\bibinfo {author} {\bibfnamefont {A.~J.}\ \bibnamefont
				{Liu}}\ and\ \bibinfo {author} {\bibfnamefont {S.~R.}\ \bibnamefont
				{Nagel}},\ }\href {\doibase 10.1038/23819} {\bibfield  {journal} {\bibinfo
				{journal} {Nature}\ }\textbf {\bibinfo {volume} {396}},\ \bibinfo {pages}
			{21} (\bibinfo {year} {1998})}\BibitemShut {NoStop}%
		\bibitem [{\citenamefont {Jaeger}(2015)}]{Jaeger2015}%
		\BibitemOpen
		\bibfield  {author} {\bibinfo {author} {\bibfnamefont {H.~M.}\ \bibnamefont
				{Jaeger}},\ }\href {\doibase 10.1039/C4SM01923G} {\bibfield  {journal}
			{\bibinfo  {journal} {Soft Matter}\ }\textbf {\bibinfo {volume} {11}},\
			\bibinfo {pages} {12} (\bibinfo {year} {2015})}\BibitemShut {NoStop}%
		\bibitem [{\citenamefont {Kapnistos}\ \emph {et~al.}(2000)\citenamefont
			{Kapnistos}, \citenamefont {Vlassopoulos}, \citenamefont {Fytas},
			\citenamefont {Mortensen}, \citenamefont {Fleischer},\ and\ \citenamefont
			{Roovers}}]{Kapnistos2000}%
		\BibitemOpen
		\bibfield  {author} {\bibinfo {author} {\bibfnamefont {M.}~\bibnamefont
				{Kapnistos}}, \bibinfo {author} {\bibfnamefont {D.}~\bibnamefont
				{Vlassopoulos}}, \bibinfo {author} {\bibfnamefont {G.}~\bibnamefont {Fytas}},
			\bibinfo {author} {\bibfnamefont {K.}~\bibnamefont {Mortensen}}, \bibinfo
			{author} {\bibfnamefont {G.}~\bibnamefont {Fleischer}}, \ and\ \bibinfo
			{author} {\bibfnamefont {J.}~\bibnamefont {Roovers}},\ }\href {\doibase
			10.1103/PhysRevLett.85.4072} {\bibfield  {journal} {\bibinfo  {journal}
				{Physical Review Letters}\ }\textbf {\bibinfo {volume} {85}},\ \bibinfo
			{pages} {4072} (\bibinfo {year} {2000})}\BibitemShut {NoStop}%
		\bibitem [{\citenamefont {Nagel}(2017)}]{Nagel2017}%
		\BibitemOpen
		\bibfield  {author} {\bibinfo {author} {\bibfnamefont {S.~R.}\ \bibnamefont
				{Nagel}},\ }\href {\doibase 10.1103/RevModPhys.89.025002} {\bibfield
			{journal} {\bibinfo  {journal} {Reviews of Modern Physics}\ }\textbf
			{\bibinfo {volume} {89}},\ \bibinfo {pages} {025002, 1} (\bibinfo {year}
			{2017})}\BibitemShut {NoStop}%
		\bibitem [{\citenamefont {Grindy}\ \emph {et~al.}(2015)\citenamefont {Grindy},
			\citenamefont {Learsch}, \citenamefont {Mozhdehi}, \citenamefont {Cheng},
			\citenamefont {Barrett}, \citenamefont {Guan}, \citenamefont {Messersmith},\
			and\ \citenamefont {Holten-Andersen}}]{Grindy2015}%
		\BibitemOpen
		\bibfield  {author} {\bibinfo {author} {\bibfnamefont {S.~C.}\ \bibnamefont
				{Grindy}}, \bibinfo {author} {\bibfnamefont {R.}~\bibnamefont {Learsch}},
			\bibinfo {author} {\bibfnamefont {D.}~\bibnamefont {Mozhdehi}}, \bibinfo
			{author} {\bibfnamefont {J.}~\bibnamefont {Cheng}}, \bibinfo {author}
			{\bibfnamefont {D.~G.}\ \bibnamefont {Barrett}}, \bibinfo {author}
			{\bibfnamefont {Z.}~\bibnamefont {Guan}}, \bibinfo {author} {\bibfnamefont
				{P.~B.}\ \bibnamefont {Messersmith}}, \ and\ \bibinfo {author} {\bibfnamefont
				{N.}~\bibnamefont {Holten-Andersen}},\ }\href {\doibase 10.1038/nmat4401}
		{\bibfield  {journal} {\bibinfo  {journal} {Nature Materials}\ }\textbf
			{\bibinfo {volume} {14}},\ \bibinfo {pages} {1210} (\bibinfo {year}
			{2015})}\BibitemShut {NoStop}%
		\bibitem [{\citenamefont {Bertoldi}\ \emph {et~al.}(2017)\citenamefont
			{Bertoldi}, \citenamefont {Vitelli}, \citenamefont {Christensen},\ and\
			\citenamefont {{Van Hecke}}}]{Bertoldi2017}%
		\BibitemOpen
		\bibfield  {author} {\bibinfo {author} {\bibfnamefont {K.}~\bibnamefont
				{Bertoldi}}, \bibinfo {author} {\bibfnamefont {V.}~\bibnamefont {Vitelli}},
			\bibinfo {author} {\bibfnamefont {J.}~\bibnamefont {Christensen}}, \ and\
			\bibinfo {author} {\bibfnamefont {M.}~\bibnamefont {{Van Hecke}}},\ }\href
		{\doibase 10.1038/natrevmats.2017.66} {\bibfield  {journal} {\bibinfo
				{journal} {Nature Reviews Materials}\ }\textbf {\bibinfo {volume} {2}},\
			\bibinfo {pages} {17066, 1} (\bibinfo {year} {2017})}\BibitemShut {NoStop}%
		\bibitem [{\citenamefont {Brown}\ \emph {et~al.}(2010)\citenamefont {Brown},
			\citenamefont {Rodenberg}, \citenamefont {Amend}, \citenamefont {Mozeika},
			\citenamefont {Steltz}, \citenamefont {Zakin}, \citenamefont {Lipson},
			\citenamefont {Jaeger},\ and\ \citenamefont {Lipson}}]{Brown2010}%
		\BibitemOpen
		\bibfield  {author} {\bibinfo {author} {\bibfnamefont {E.}~\bibnamefont
				{Brown}}, \bibinfo {author} {\bibfnamefont {N.}~\bibnamefont {Rodenberg}},
			\bibinfo {author} {\bibfnamefont {J.}~\bibnamefont {Amend}}, \bibinfo
			{author} {\bibfnamefont {A.}~\bibnamefont {Mozeika}}, \bibinfo {author}
			{\bibfnamefont {E.}~\bibnamefont {Steltz}}, \bibinfo {author} {\bibfnamefont
				{M.~R.}\ \bibnamefont {Zakin}}, \bibinfo {author} {\bibfnamefont
				{H.}~\bibnamefont {Lipson}}, \bibinfo {author} {\bibfnamefont {H.~M.}\
				\bibnamefont {Jaeger}}, \ and\ \bibinfo {author} {\bibfnamefont
				{H.}~\bibnamefont {Lipson}},\ }\href {\doibase 10.1109/TRO.2011.2171093}
		{\bibfield  {journal} {\bibinfo  {journal} {IEEE Transactions on Robotics}\
			}\textbf {\bibinfo {volume} {107}},\ \bibinfo {pages} {18809} (\bibinfo
			{year} {2010})}\BibitemShut {NoStop}%
		\bibitem [{\citenamefont {Bird}\ \emph {et~al.}(1987)\citenamefont {Bird},
			\citenamefont {Armstrong},\ and\ \citenamefont {Hassager}}]{Bird1987}%
		\BibitemOpen
		\bibfield  {author} {\bibinfo {author} {\bibfnamefont {R.~B.}\ \bibnamefont
				{Bird}}, \bibinfo {author} {\bibfnamefont {R.~C.}\ \bibnamefont {Armstrong}},
			\ and\ \bibinfo {author} {\bibfnamefont {O.}~\bibnamefont {Hassager}},\
		}\href {http://www.osti.gov/scitech/biblio/6164599} {\emph {\bibinfo {title}
				{{Dynamics of Polymeric Liquids. Vol. 1: Fluid mechanics}}}},\ \bibinfo
		{edition} {2nd}\ ed.\ (\bibinfo  {publisher} {John Wiley and Sons Inc.,New
			York, NY},\ \bibinfo {year} {1987})\BibitemShut {NoStop}%
		\bibitem [{\citenamefont {Mours}\ and\ \citenamefont
			{Winter}(1994)}]{Mours1994}%
		\BibitemOpen
		\bibfield  {author} {\bibinfo {author} {\bibfnamefont {M.}~\bibnamefont
				{Mours}}\ and\ \bibinfo {author} {\bibfnamefont {H.~H.}\ \bibnamefont
				{Winter}},\ }\href {\doibase 10.1007/BF00366581} {\bibfield  {journal}
			{\bibinfo  {journal} {Rheologica Acta}\ }\textbf {\bibinfo {volume} {33}},\
			\bibinfo {pages} {385} (\bibinfo {year} {1994})}\BibitemShut {NoStop}%
		\bibitem [{\citenamefont {Adam}\ \emph {et~al.}(1988)\citenamefont {Adam},
			\citenamefont {Delsanti}, \citenamefont {Munch},\ and\ \citenamefont
			{Durand}}]{Adam1988}%
		\BibitemOpen
		\bibfield  {author} {\bibinfo {author} {\bibfnamefont {M.}~\bibnamefont
				{Adam}}, \bibinfo {author} {\bibfnamefont {M.}~\bibnamefont {Delsanti}},
			\bibinfo {author} {\bibfnamefont {J.~P.}\ \bibnamefont {Munch}}, \ and\
			\bibinfo {author} {\bibfnamefont {D.}~\bibnamefont {Durand}},\ }\href
		{\doibase https://doi.org/10.1103/PhysRevLett.61.706} {\bibfield  {journal}
			{\bibinfo  {journal} {Physical Review Letters}\ }\textbf {\bibinfo {volume}
				{61}},\ \bibinfo {pages} {706} (\bibinfo {year} {1988})}\BibitemShut
		{NoStop}%
		\bibitem [{\citenamefont {Kroon}\ \emph {et~al.}(1996)\citenamefont {Kroon},
			\citenamefont {Wegdam},\ and\ \citenamefont {Sprik}}]{Kroon1996}%
		\BibitemOpen
		\bibfield  {author} {\bibinfo {author} {\bibfnamefont {M.}~\bibnamefont
				{Kroon}}, \bibinfo {author} {\bibfnamefont {G.~H.}\ \bibnamefont {Wegdam}}, \
			and\ \bibinfo {author} {\bibfnamefont {R.}~\bibnamefont {Sprik}},\ }\href
		{\doibase https://doi.org/10.1103/PhysRevE.54.6541} {\bibfield  {journal}
			{\bibinfo  {journal} {Physical Review E}\ }\textbf {\bibinfo {volume} {54}},\
			\bibinfo {pages} {6541} (\bibinfo {year} {1996})}\BibitemShut {NoStop}%
		\bibitem [{\citenamefont {Romer}\ \emph {et~al.}(2000)\citenamefont {Romer},
			\citenamefont {Scheffold},\ and\ \citenamefont {Schurtenberger}}]{Romer2000}%
		\BibitemOpen
		\bibfield  {author} {\bibinfo {author} {\bibfnamefont {S.}~\bibnamefont
				{Romer}}, \bibinfo {author} {\bibfnamefont {F.}~\bibnamefont {Scheffold}}, \
			and\ \bibinfo {author} {\bibfnamefont {P.}~\bibnamefont {Schurtenberger}},\
		}\href {\doibase 10.1103/PhysRevLett.85.4980} {\bibfield  {journal} {\bibinfo
				{journal} {Physical Review Letters}\ }\textbf {\bibinfo {volume} {85}},\
			\bibinfo {pages} {4980} (\bibinfo {year} {2000})}\BibitemShut {NoStop}%
		\bibitem [{\citenamefont {Duri}\ \emph {et~al.}(2005)\citenamefont {Duri},
			\citenamefont {Bissig}, \citenamefont {Trappe},\ and\ \citenamefont
			{Cipelletti}}]{Duri2005}%
		\BibitemOpen
		\bibfield  {author} {\bibinfo {author} {\bibfnamefont {A.}~\bibnamefont
				{Duri}}, \bibinfo {author} {\bibfnamefont {H.}~\bibnamefont {Bissig}},
			\bibinfo {author} {\bibfnamefont {V.}~\bibnamefont {Trappe}}, \ and\ \bibinfo
			{author} {\bibfnamefont {L.}~\bibnamefont {Cipelletti}},\ }\href {\doibase
			10.1103/PhysRevE.72.051401} {\bibfield  {journal} {\bibinfo  {journal}
				{Physical Review E}\ }\textbf {\bibinfo {volume} {72}},\ \bibinfo {pages}
			{051401, 1} (\bibinfo {year} {2005})}\BibitemShut {NoStop}%
		\bibitem [{\citenamefont {Eberle}\ \emph {et~al.}(2011)\citenamefont {Eberle},
			\citenamefont {Wagner},\ and\ \citenamefont
			{Casta{\~{n}}eda-Priego}}]{Eberle2011}%
		\BibitemOpen
		\bibfield  {author} {\bibinfo {author} {\bibfnamefont {A.~P.~R.}\
				\bibnamefont {Eberle}}, \bibinfo {author} {\bibfnamefont {N.~J.}\
				\bibnamefont {Wagner}}, \ and\ \bibinfo {author} {\bibfnamefont
				{R.}~\bibnamefont {Casta{\~{n}}eda-Priego}},\ }\href {\doibase
			10.1103/PhysRevLett.106.105704} {\bibfield  {journal} {\bibinfo  {journal}
				{Physical Review Letters}\ }\textbf {\bibinfo {volume} {106}},\ \bibinfo
			{pages} {105704, 1} (\bibinfo {year} {2011})}\BibitemShut {NoStop}%
		\bibitem [{\citenamefont {Weeks}\ \emph {et~al.}(2000)\citenamefont {Weeks},
			\citenamefont {Crocker}, \citenamefont {Levitt}, \citenamefont {Schofield},\
			and\ \citenamefont {Weitz}}]{Weeks2000}%
		\BibitemOpen
		\bibfield  {author} {\bibinfo {author} {\bibfnamefont {E.~R.}\ \bibnamefont
				{Weeks}}, \bibinfo {author} {\bibfnamefont {J.}~\bibnamefont {Crocker}},
			\bibinfo {author} {\bibfnamefont {A.~C.}\ \bibnamefont {Levitt}}, \bibinfo
			{author} {\bibfnamefont {A.}~\bibnamefont {Schofield}}, \ and\ \bibinfo
			{author} {\bibfnamefont {D.~A.}\ \bibnamefont {Weitz}},\ }\href {\doibase
			10.1126/science.287.5453.627} {\bibfield  {journal} {\bibinfo  {journal}
				{Science}\ }\textbf {\bibinfo {volume} {287}},\ \bibinfo {pages} {627}
			(\bibinfo {year} {2000})}\BibitemShut {NoStop}%
		\bibitem [{\citenamefont {Assink}\ and\ \citenamefont
			{Kay}(1991)}]{Assink1991}%
		\BibitemOpen
		\bibfield  {author} {\bibinfo {author} {\bibfnamefont {R.~A.}\ \bibnamefont
				{Assink}}\ and\ \bibinfo {author} {\bibfnamefont {B.~D.}\ \bibnamefont
				{Kay}},\ }\href {\doibase 10.1146/annurev.ms.21.080191.002423} {\bibfield
			{journal} {\bibinfo  {journal} {Annual Review of Material Science}\ }\textbf
			{\bibinfo {volume} {21}},\ \bibinfo {pages} {491} (\bibinfo {year}
			{1991})}\BibitemShut {NoStop}%
		\bibitem [{\citenamefont {Bonn}\ \emph {et~al.}(2008)\citenamefont {Bonn},
			\citenamefont {Rodts}, \citenamefont {Groenink}, \citenamefont {Rafa{\"{i}}},
			\citenamefont {Shahidzadeh-Bonn},\ and\ \citenamefont {Coussot}}]{Bonn2008}%
		\BibitemOpen
		\bibfield  {author} {\bibinfo {author} {\bibfnamefont {D.}~\bibnamefont
				{Bonn}}, \bibinfo {author} {\bibfnamefont {S.}~\bibnamefont {Rodts}},
			\bibinfo {author} {\bibfnamefont {M.}~\bibnamefont {Groenink}}, \bibinfo
			{author} {\bibfnamefont {S.}~\bibnamefont {Rafa{\"{i}}}}, \bibinfo {author}
			{\bibfnamefont {N.}~\bibnamefont {Shahidzadeh-Bonn}}, \ and\ \bibinfo
			{author} {\bibfnamefont {P.}~\bibnamefont {Coussot}},\ }\href {\doibase
			10.1146/annurev.fluid.40.111406.102211} {\bibfield  {journal} {\bibinfo
				{journal} {Annual Review of Fluid Mechanics}\ }\textbf {\bibinfo {volume}
				{40}},\ \bibinfo {pages} {209} (\bibinfo {year} {2008})}\BibitemShut
		{NoStop}%
		\bibitem [{\citenamefont {Larsen}\ and\ \citenamefont
			{Furst}(2008)}]{Larsen2008}%
		\BibitemOpen
		\bibfield  {author} {\bibinfo {author} {\bibfnamefont {T.~H.}\ \bibnamefont
				{Larsen}}\ and\ \bibinfo {author} {\bibfnamefont {E.~M.}\ \bibnamefont
				{Furst}},\ }\href {\doibase 10.1103/PhysRevLett.100.146001} {\bibfield
			{journal} {\bibinfo  {journal} {Physical Review Letters}\ }\textbf {\bibinfo
				{volume} {100}},\ \bibinfo {pages} {1} (\bibinfo {year} {2008})}\BibitemShut
		{NoStop}%
		\bibitem [{\citenamefont {Edera}\ \emph {et~al.}(2017)\citenamefont {Edera},
			\citenamefont {Bergamini}, \citenamefont {Trappe}, \citenamefont {Giavazzi},\
			and\ \citenamefont {Cerbino}}]{Edera2017}%
		\BibitemOpen
		\bibfield  {author} {\bibinfo {author} {\bibfnamefont {P.}~\bibnamefont
				{Edera}}, \bibinfo {author} {\bibfnamefont {D.}~\bibnamefont {Bergamini}},
			\bibinfo {author} {\bibfnamefont {V.}~\bibnamefont {Trappe}}, \bibinfo
			{author} {\bibfnamefont {F.}~\bibnamefont {Giavazzi}}, \ and\ \bibinfo
			{author} {\bibfnamefont {R.}~\bibnamefont {Cerbino}},\ }\href {\doibase
			10.1103/PhysRevMaterials.1.073804} {\bibfield  {journal} {\bibinfo  {journal}
				{Physical Review Materials}\ }\textbf {\bibinfo {volume} {1}},\ \bibinfo
			{pages} {073804 1} (\bibinfo {year} {2017})}\BibitemShut {NoStop}%
		\bibitem [{\citenamefont {Nia}\ \emph {et~al.}(2013)\citenamefont {Nia},
			\citenamefont {Bozchalooi}, \citenamefont {Li}, \citenamefont {Han},
			\citenamefont {Hung}, \citenamefont {Frank}, \citenamefont {Youcef-Toumi},
			\citenamefont {Ortiz},\ and\ \citenamefont {Grodzinsky}}]{Nia2013}%
		\BibitemOpen
		\bibfield  {author} {\bibinfo {author} {\bibfnamefont {H.~T.}\ \bibnamefont
				{Nia}}, \bibinfo {author} {\bibfnamefont {I.~S.}\ \bibnamefont {Bozchalooi}},
			\bibinfo {author} {\bibfnamefont {Y.}~\bibnamefont {Li}}, \bibinfo {author}
			{\bibfnamefont {L.}~\bibnamefont {Han}}, \bibinfo {author} {\bibfnamefont
				{H.~H.}\ \bibnamefont {Hung}}, \bibinfo {author} {\bibfnamefont
				{E.}~\bibnamefont {Frank}}, \bibinfo {author} {\bibfnamefont
				{K.}~\bibnamefont {Youcef-Toumi}}, \bibinfo {author} {\bibfnamefont
				{C.}~\bibnamefont {Ortiz}}, \ and\ \bibinfo {author} {\bibfnamefont
				{A.}~\bibnamefont {Grodzinsky}},\ }\href {\doibase 10.1016/j.bpj.2013.02.048}
		{\bibfield  {journal} {\bibinfo  {journal} {Biophysical Journal}\ }\textbf
			{\bibinfo {volume} {104}},\ \bibinfo {pages} {1529} (\bibinfo {year}
			{2013})}\BibitemShut {NoStop}%
		\bibitem [{\citenamefont {Menard}(1999)}]{Menard1999}%
		\BibitemOpen
		\bibfield  {author} {\bibinfo {author} {\bibfnamefont {K.~P.}\ \bibnamefont
				{Menard}},\ }\href
		{https://books.google.com/books?id=AMCjRtKY17IC{\&}pg=PA1{\&}lpg=PA1{\&}dq=first+to+perform+oscillatory+measurements{\&}source=bl{\&}ots=S-ND3aD9M5{\&}sig=8ewniEePvvkGRjBhnUPgiWSZvD8{\&}hl=en{\&}sa=X{\&}ved=0ahUKEwivzuPdvLLZAhVs9IMKHeyaByYQ6AEIbzAF{\#}v=onepage{\&}q=first
			to perform oscil} {\emph {\bibinfo {title} {{Dynamic Mechanical Analysis: A
						Practical introduction}}}}\ (\bibinfo  {publisher} {CRC Press},\ \bibinfo
		{year} {1999})\BibitemShut {NoStop}%
		\bibitem [{\citenamefont {Fausti}\ and\ \citenamefont
			{Farina}(2000)}]{Fausti2000}%
		\BibitemOpen
		\bibfield  {author} {\bibinfo {author} {\bibfnamefont {P.}~\bibnamefont
				{Fausti}}\ and\ \bibinfo {author} {\bibfnamefont {A.}~\bibnamefont
				{Farina}},\ }\href {\doibase 10.1006/jsvi.1999.2694} {\bibfield  {journal}
			{\bibinfo  {journal} {Journal of Sound and Vibration}\ }\textbf {\bibinfo
				{volume} {232}},\ \bibinfo {pages} {213} (\bibinfo {year}
			{2000})}\BibitemShut {NoStop}%
		\bibitem [{\citenamefont {Klauder}\ \emph {et~al.}(1960)\citenamefont
			{Klauder}, \citenamefont {Price}, \citenamefont {Darlington},\ and\
			\citenamefont {Albersheim}}]{Klauder1960}%
		\BibitemOpen
		\bibfield  {author} {\bibinfo {author} {\bibfnamefont {J.}~\bibnamefont
				{Klauder}}, \bibinfo {author} {\bibfnamefont {A.}~\bibnamefont {Price}},
			\bibinfo {author} {\bibfnamefont {S.}~\bibnamefont {Darlington}}, \ and\
			\bibinfo {author} {\bibfnamefont {W.}~\bibnamefont {Albersheim}},\
		}\href@noop {} {\bibfield  {journal} {\bibinfo  {journal} {The Bell System
					Technical Journal}\ }\textbf {\bibinfo {volume} {39}},\ \bibinfo {pages}
			{745} (\bibinfo {year} {1960})}\BibitemShut {NoStop}%
		\bibitem [{\citenamefont {Holly}\ \emph {et~al.}(1988)\citenamefont {Holly},
			\citenamefont {Venkataraman}, \citenamefont {Chambon},\ and\ \citenamefont
			{Winter}}]{Holly1988}%
		\BibitemOpen
		\bibfield  {author} {\bibinfo {author} {\bibfnamefont {E.~E.}\ \bibnamefont
				{Holly}}, \bibinfo {author} {\bibfnamefont {S.~K.}\ \bibnamefont
				{Venkataraman}}, \bibinfo {author} {\bibfnamefont {F.}~\bibnamefont
				{Chambon}}, \ and\ \bibinfo {author} {\bibfnamefont {H.~H.}\ \bibnamefont
				{Winter}},\ }\href {\doibase 10.1016/0377-0257(88)80002-8} {\bibfield
			{journal} {\bibinfo  {journal} {Journal of Non-Newtonian Fluid Mechanics}\
			}\textbf {\bibinfo {volume} {27}},\ \bibinfo {pages} {17} (\bibinfo {year}
			{1988})}\BibitemShut {NoStop}%
		\bibitem [{\citenamefont {M{\"{u}}ller}\ and\ \citenamefont
			{Massarani}(2001)}]{Muller2001}%
		\BibitemOpen
		\bibfield  {author} {\bibinfo {author} {\bibfnamefont {S.}~\bibnamefont
				{M{\"{u}}ller}}\ and\ \bibinfo {author} {\bibfnamefont {P.}~\bibnamefont
				{Massarani}},\ }\href@noop {} {\bibfield  {journal} {\bibinfo  {journal}
				{Journal of the Audio Engineering Society}\ }\textbf {\bibinfo {volume}
				{49}},\ \bibinfo {pages} {443} (\bibinfo {year} {2001})}\BibitemShut
		{NoStop}%
		\bibitem [{\citenamefont {Field}\ \emph {et~al.}(1996)\citenamefont {Field},
			\citenamefont {Swain},\ and\ \citenamefont {Phan-Thien}}]{Field1996}%
		\BibitemOpen
		\bibfield  {author} {\bibinfo {author} {\bibfnamefont {J.}~\bibnamefont
				{Field}}, \bibinfo {author} {\bibfnamefont {M.}~\bibnamefont {Swain}}, \ and\
			\bibinfo {author} {\bibfnamefont {N.}~\bibnamefont {Phan-Thien}},\ }\href
		{\doibase https://doi.org/10.1016/0377-0257(96)01445-0} {\bibfield  {journal}
			{\bibinfo  {journal} {Journal of Non-Newtonian Fluid Mechanics}\ }\textbf
			{\bibinfo {volume} {65}},\ \bibinfo {pages} {177} (\bibinfo {year}
			{1996})}\BibitemShut {NoStop}%
		\bibitem [{\citenamefont {Tassieri}\ \emph {et~al.}(2016)\citenamefont
			{Tassieri}, \citenamefont {Laurati}, \citenamefont {Curtis}, \citenamefont
			{Auhl}, \citenamefont {Coppola}, \citenamefont {Scalfati}, \citenamefont
			{Williams}, \citenamefont {Cooper}, \citenamefont {Laurati}, \citenamefont
			{Curtis}, \citenamefont {Scalfati}, \citenamefont {Hawkins},\ and\
			\citenamefont {Cooper}}]{Tassieri2016}%
		\BibitemOpen
		\bibfield  {author} {\bibinfo {author} {\bibfnamefont {M.}~\bibnamefont
				{Tassieri}}, \bibinfo {author} {\bibfnamefont {M.}~\bibnamefont {Laurati}},
			\bibinfo {author} {\bibfnamefont {D.~J.}\ \bibnamefont {Curtis}}, \bibinfo
			{author} {\bibfnamefont {D.~W.}\ \bibnamefont {Auhl}}, \bibinfo {author}
			{\bibfnamefont {S.}~\bibnamefont {Coppola}}, \bibinfo {author} {\bibfnamefont
				{A.}~\bibnamefont {Scalfati}}, \bibinfo {author} {\bibfnamefont {P.~R.}\
				\bibnamefont {Williams}}, \bibinfo {author} {\bibfnamefont {J.~M.}\
				\bibnamefont {Cooper}}, \bibinfo {author} {\bibfnamefont {M.}~\bibnamefont
				{Laurati}}, \bibinfo {author} {\bibfnamefont {D.~J.}\ \bibnamefont {Curtis}},
			\bibinfo {author} {\bibfnamefont {A.}~\bibnamefont {Scalfati}}, \bibinfo
			{author} {\bibfnamefont {K.}~\bibnamefont {Hawkins}}, \ and\ \bibinfo
			{author} {\bibfnamefont {J.~M.}\ \bibnamefont {Cooper}},\ }\href {\doibase
			10.1122/1.4953443} {\bibfield  {journal} {\bibinfo  {journal} {Journal of
					Rheology}\ }\textbf {\bibinfo {volume} {60}},\ \bibinfo {pages} {649}
			(\bibinfo {year} {2016})}\BibitemShut {NoStop}%
		\bibitem [{\citenamefont {Schock}\ \emph {et~al.}(1989)\citenamefont {Schock},
			\citenamefont {Leblanc},\ and\ \citenamefont {Mayer}}]{Schock1989}%
		\BibitemOpen
		\bibfield  {author} {\bibinfo {author} {\bibfnamefont {S.~G.}\ \bibnamefont
				{Schock}}, \bibinfo {author} {\bibfnamefont {L.~R.}\ \bibnamefont {Leblanc}},
			\ and\ \bibinfo {author} {\bibfnamefont {L.~A.}\ \bibnamefont {Mayer}},\
		}\href {\doibase 10.1190/1.1442670} {\bibfield  {journal} {\bibinfo
				{journal} {Geophysics}\ }\textbf {\bibinfo {volume} {54}},\ \bibinfo {pages}
			{445} (\bibinfo {year} {1989})}\BibitemShut {NoStop}%
		\bibitem [{\citenamefont {Heyes}\ \emph {et~al.}(1994)\citenamefont {Heyes},
			\citenamefont {Mitchell}, \citenamefont {Visschert},\ and\ \citenamefont
			{Melrose}}]{Heyes1994}%
		\BibitemOpen
		\bibfield  {author} {\bibinfo {author} {\bibfnamefont {M.}~\bibnamefont
				{Heyes}}, \bibinfo {author} {\bibfnamefont {P.}~\bibnamefont {Mitchell}},
			\bibinfo {author} {\bibfnamefont {P.}~\bibnamefont {Visschert}}, \ and\
			\bibinfo {author} {\bibfnamefont {J.}~\bibnamefont {Melrose}},\ }\href
		{\doibase 10.1039/FT9949001133} {\bibfield  {journal} {\bibinfo  {journal}
				{Journal of the Chemical Society Faraday Transactions}\ }\textbf {\bibinfo
				{volume} {90}},\ \bibinfo {pages} {1133} (\bibinfo {year}
			{1994})}\BibitemShut {NoStop}%
		\bibitem [{\citenamefont {Ghiringhelli}\ \emph {et~al.}(2012)\citenamefont
			{Ghiringhelli}, \citenamefont {Roux}, \citenamefont {Bleses}, \citenamefont
			{Galliard},\ and\ \citenamefont {Caton}}]{Ghiringhelli2012}%
		\BibitemOpen
		\bibfield  {author} {\bibinfo {author} {\bibfnamefont {E.}~\bibnamefont
				{Ghiringhelli}}, \bibinfo {author} {\bibfnamefont {D.}~\bibnamefont {Roux}},
			\bibinfo {author} {\bibfnamefont {D.}~\bibnamefont {Bleses}}, \bibinfo
			{author} {\bibfnamefont {H.}~\bibnamefont {Galliard}}, \ and\ \bibinfo
			{author} {\bibfnamefont {F.}~\bibnamefont {Caton}},\ }\href {\doibase
			10.1007/s00397-012-0616-z} {\bibfield  {journal} {\bibinfo  {journal}
				{Rheologica Acta}\ }\textbf {\bibinfo {volume} {51}},\ \bibinfo {pages} {413}
			(\bibinfo {year} {2012})}\BibitemShut {NoStop}%
		\bibitem [{\citenamefont {Curtis}\ \emph {et~al.}(2014)\citenamefont {Curtis},
			\citenamefont {Holder}, \citenamefont {Badiei}, \citenamefont {Claypole},
			\citenamefont {Walters}, \citenamefont {Thomas}, \citenamefont {Barrow},
			\citenamefont {Deganello}, \citenamefont {Brown}, \citenamefont {Williams},\
			and\ \citenamefont {Hawkins}}]{Curtis2014}%
		\BibitemOpen
		\bibfield  {author} {\bibinfo {author} {\bibfnamefont {D.~J.}\ \bibnamefont
				{Curtis}}, \bibinfo {author} {\bibfnamefont {A.}~\bibnamefont {Holder}},
			\bibinfo {author} {\bibfnamefont {N.}~\bibnamefont {Badiei}}, \bibinfo
			{author} {\bibfnamefont {J.}~\bibnamefont {Claypole}}, \bibinfo {author}
			{\bibfnamefont {M.}~\bibnamefont {Walters}}, \bibinfo {author} {\bibfnamefont
				{B.}~\bibnamefont {Thomas}}, \bibinfo {author} {\bibfnamefont
				{M.}~\bibnamefont {Barrow}}, \bibinfo {author} {\bibfnamefont
				{D.}~\bibnamefont {Deganello}}, \bibinfo {author} {\bibfnamefont {M.~R.}\
				\bibnamefont {Brown}}, \bibinfo {author} {\bibfnamefont {P.~R.}\ \bibnamefont
				{Williams}}, \ and\ \bibinfo {author} {\bibfnamefont {K.}~\bibnamefont
				{Hawkins}},\ }\href {\doibase 10.1016/j.jnnfm.2015.01.003} {\bibfield
			{journal} {\bibinfo  {journal} {Journal of Non-Newtonian Fluid Mechanics}\
			}\textbf {\bibinfo {volume} {222}},\ \bibinfo {pages} {253} (\bibinfo {year}
			{2014})}\BibitemShut {NoStop}%
		\bibitem [{\citenamefont {Rouyer}\ and\ \citenamefont
			{Poulesquen}(2015)}]{Rouyer2015}%
		\BibitemOpen
		\bibfield  {author} {\bibinfo {author} {\bibfnamefont {J.}~\bibnamefont
				{Rouyer}}\ and\ \bibinfo {author} {\bibfnamefont {A.}~\bibnamefont
				{Poulesquen}},\ }\href {\doibase 10.1111/jace.13480} {\bibfield  {journal}
			{\bibinfo  {journal} {Journal of the American Ceramic Society}\ }\textbf
			{\bibinfo {volume} {98}},\ \bibinfo {pages} {1580} (\bibinfo {year}
			{2015})}\BibitemShut {NoStop}%
		\bibitem [{\citenamefont {Kowatsch}\ and\ \citenamefont
			{Stocker}(1982)}]{Kowatsch1982}%
		\BibitemOpen
		\bibfield  {author} {\bibinfo {author} {\bibfnamefont {M.}~\bibnamefont
				{Kowatsch}}\ and\ \bibinfo {author} {\bibfnamefont {H.}~\bibnamefont
				{Stocker}},\ }\href@noop {} {\bibfield  {journal} {\bibinfo  {journal}
				{Communications, Radar and Signal Processing, IEE Proceedings}\ }\textbf
			{\bibinfo {volume} {129}},\ \bibinfo {pages} {41} (\bibinfo {year}
			{1982})}\BibitemShut {NoStop}%
		\bibitem [{\citenamefont {M{\"{u}}ller}(2008)}]{Muller2008}%
		\BibitemOpen
		\bibfield  {author} {\bibinfo {author} {\bibfnamefont {S.}~\bibnamefont
				{M{\"{u}}ller}},\ }in\ \href {\doibase 10.1007/978-0-387-30441-0_5} {\emph
			{\bibinfo {booktitle} {Handbook of Signal Processing in Acoustics}}},\
		\bibinfo {editor} {edited by\ \bibinfo {editor} {\bibfnamefont
				{D.}~\bibnamefont {Havelock}}, \bibinfo {editor} {\bibfnamefont
				{S.}~\bibnamefont {Kuwano}}, \ and\ \bibinfo {editor} {\bibfnamefont
				{M.}~\bibnamefont {Vorl{\"{a}}nder}}}\ (\bibinfo  {publisher} {Springer New
			York},\ \bibinfo {year} {2008})\ pp.\ \bibinfo {pages} {65--85}\BibitemShut
		{NoStop}%
		\bibitem [{\citenamefont {Pintelon}\ and\ \citenamefont
			{Schoukens}(2012)}]{Pintelon2012}%
		\BibitemOpen
		\bibfield  {author} {\bibinfo {author} {\bibfnamefont {R.}~\bibnamefont
				{Pintelon}}\ and\ \bibinfo {author} {\bibfnamefont {J.}~\bibnamefont
				{Schoukens}},\ }\href@noop {} {\emph {\bibinfo {title} {{System
						Identification: A Frequency Domain Approach}}}}\ (\bibinfo  {publisher}
		{Wiley-IEEE Press},\ \bibinfo {year} {2012})\ p.\ \bibinfo {pages}
		{788}\BibitemShut {NoStop}%
		\bibitem [{\citenamefont {Oppenheim}\ \emph {et~al.}(1983)\citenamefont
			{Oppenheim}, \citenamefont {Willsky},\ and\ \citenamefont
			{Young}}]{Oppenheim1983}%
		\BibitemOpen
		\bibfield  {author} {\bibinfo {author} {\bibfnamefont {A.~V.}\ \bibnamefont
				{Oppenheim}}, \bibinfo {author} {\bibfnamefont {A.~S.}\ \bibnamefont
				{Willsky}}, \ and\ \bibinfo {author} {\bibfnamefont {I.~T.}\ \bibnamefont
				{Young}},\ }\href@noop {} {\emph {\bibinfo {title} {{Signals and Systems}}}}\
		(\bibinfo  {publisher} {Prentiece-Hall, Inc.},\ \bibinfo {year} {1983})\ p.\
		\bibinfo {pages} {796}\BibitemShut {NoStop}%
		\bibitem [{\citenamefont {Stowell}\ and\ \citenamefont
			{Plumbley}(2014)}]{Stowell2014}%
		\BibitemOpen
		\bibfield  {author} {\bibinfo {author} {\bibfnamefont {D.}~\bibnamefont
				{Stowell}}\ and\ \bibinfo {author} {\bibfnamefont {M.~D.}\ \bibnamefont
				{Plumbley}},\ }\href {\doibase 10.1111/2041-210X.12223} {\bibfield  {journal}
			{\bibinfo  {journal} {Methods in Ecology and Evolution}\ }\textbf {\bibinfo
				{volume} {5}},\ \bibinfo {pages} {901} (\bibinfo {year} {2014})}\BibitemShut
		{NoStop}%
		\bibitem [{\citenamefont {Abbott}\ \emph {et~al.}(2016)\citenamefont {Abbott},
			\citenamefont {Abbott}, \citenamefont {Abbott}, \citenamefont {Abernathy},
			\citenamefont {Acernese},\ and\ \citenamefont {Ackley}}]{Abbott2016a}%
		\BibitemOpen
		\bibfield  {author} {\bibinfo {author} {\bibfnamefont {B.~P.}\ \bibnamefont
				{Abbott}}, \bibinfo {author} {\bibfnamefont {R.}~\bibnamefont {Abbott}},
			\bibinfo {author} {\bibfnamefont {T.~D.}\ \bibnamefont {Abbott}}, \bibinfo
			{author} {\bibfnamefont {M.~R.}\ \bibnamefont {Abernathy}}, \bibinfo {author}
			{\bibfnamefont {F.}~\bibnamefont {Acernese}}, \ and\ \bibinfo {author}
			{\bibfnamefont {K.}~\bibnamefont {Ackley}},\ }\href {\doibase
			10.1103/PhysRevLett.116.061102} {\bibfield  {journal} {\bibinfo  {journal}
				{Phys. Rev. Lett.}\ }\textbf {\bibinfo {volume} {116}},\ \bibinfo {pages}
			{61102} (\bibinfo {year} {2016})}\BibitemShut {NoStop}%
		\bibitem [{\citenamefont {Au}\ and\ \citenamefont {Simmons}(2007)}]{Au2007}%
		\BibitemOpen
		\bibfield  {author} {\bibinfo {author} {\bibfnamefont {W.~W.~L.}\
				\bibnamefont {Au}}\ and\ \bibinfo {author} {\bibfnamefont {J.~A.}\
				\bibnamefont {Simmons}},\ }\href {\doibase 10.1063/1.2784683} {\bibfield
			{journal} {\bibinfo  {journal} {Physics Today}\ }\textbf {\bibinfo {volume}
				{60}},\ \bibinfo {pages} {40} (\bibinfo {year} {2007})}\BibitemShut {NoStop}%
		\bibitem [{\citenamefont {Madsen}\ and\ \citenamefont
			{Surlykke}(2013)}]{Madsen2013}%
		\BibitemOpen
		\bibfield  {author} {\bibinfo {author} {\bibfnamefont {P.}~\bibnamefont
				{Madsen}}\ and\ \bibinfo {author} {\bibfnamefont {A.}~\bibnamefont
				{Surlykke}},\ }\href {\doibase 10.1152/physiol.00008.2013} {\bibfield
			{journal} {\bibinfo  {journal} {Physiology}\ }\textbf {\bibinfo {volume}
				{28}},\ \bibinfo {pages} {276} (\bibinfo {year} {2013})}\BibitemShut
		{NoStop}%
		\bibitem [{\citenamefont {Isermann}\ and\ \citenamefont
			{M{\"{u}}nchhof}(2011)}]{Isermann2011}%
		\BibitemOpen
		\bibfield  {author} {\bibinfo {author} {\bibfnamefont {R.}~\bibnamefont
				{Isermann}}\ and\ \bibinfo {author} {\bibfnamefont {M.}~\bibnamefont
				{M{\"{u}}nchhof}},\ }\href {\doibase 10.1007/978-3-540-78879-9} {\emph
			{\bibinfo {title} {{Identification of Dynamic Systems}}}}\ (\bibinfo
		{publisher} {Springer-Verlag Berlin Heidelberg},\ \bibinfo {year} {2011})\
		p.\ \bibinfo {pages} {711}\BibitemShut {NoStop}%
		\bibitem [{\citenamefont {Misaridis}\ and\ \citenamefont
			{Jensen}(2005{\natexlab{a}})}]{Misaridis2005a}%
		\BibitemOpen
		\bibfield  {author} {\bibinfo {author} {\bibfnamefont {T.}~\bibnamefont
				{Misaridis}}\ and\ \bibinfo {author} {\bibfnamefont {J.~A.}\ \bibnamefont
				{Jensen}},\ }\href {\doibase 10.1109/TUFFC.2005.1406545} {\bibfield
			{journal} {\bibinfo  {journal} {IEEE Transactions on ultrasonics,
					ferroelectrics, and frequency control}\ }\textbf {\bibinfo {volume} {52}},\
			\bibinfo {pages} {177} (\bibinfo {year} {2005}{\natexlab{a}})}\BibitemShut
		{NoStop}%
		\bibitem [{\citenamefont {Misaridis}\ and\ \citenamefont
			{Jensen}(2005{\natexlab{b}})}]{Misaridis2005b}%
		\BibitemOpen
		\bibfield  {author} {\bibinfo {author} {\bibfnamefont {T.}~\bibnamefont
				{Misaridis}}\ and\ \bibinfo {author} {\bibfnamefont {J.~A.}\ \bibnamefont
				{Jensen}},\ }\href {\doibase 10.1109/TUFFC.2005.1406546} {\bibfield
			{journal} {\bibinfo  {journal} {IEEE Transactions on ultrasonics,
					ferroelectrics, and frequency control}\ }\textbf {\bibinfo {volume} {52}},\
			\bibinfo {pages} {192} (\bibinfo {year} {2005}{\natexlab{b}})}\BibitemShut
		{NoStop}%
		\bibitem [{\citenamefont {Winter}\ \emph {et~al.}(1988)\citenamefont {Winter},
			\citenamefont {Morganelli},\ and\ \citenamefont {Chambon}}]{Winter1988}%
		\BibitemOpen
		\bibfield  {author} {\bibinfo {author} {\bibfnamefont {H.~H.}\ \bibnamefont
				{Winter}}, \bibinfo {author} {\bibfnamefont {P.}~\bibnamefont {Morganelli}},
			\ and\ \bibinfo {author} {\bibfnamefont {F.}~\bibnamefont {Chambon}},\ }\href
		{\doibase 10.1021/ma00180a048} {\bibfield  {journal} {\bibinfo  {journal}
				{Macromolecules}\ }\textbf {\bibinfo {volume} {21}},\ \bibinfo {pages} {535}
			(\bibinfo {year} {1988})}\BibitemShut {NoStop}%
		\bibitem [{\citenamefont {Priestley}(1981)}]{Priestley1981}%
		\BibitemOpen
		\bibfield  {author} {\bibinfo {author} {\bibfnamefont {M.}~\bibnamefont
				{Priestley}},\ }\href@noop {} {\emph {\bibinfo {title} {{Spectral Analysis
						and Time Series}}}}\ (\bibinfo  {publisher} {Academic Press},\ \bibinfo
		{year} {1981})\ p.\ \bibinfo {pages} {890}\BibitemShut {NoStop}%
		\bibitem [{\citenamefont {Jones}(2005)}]{Jones2005}%
		\BibitemOpen
		\bibfield  {author} {\bibinfo {author} {\bibfnamefont {G.}~\bibnamefont
				{Jones}},\ }\href {\doibase http://dx.doi.org/10.1016/j.cub.2005.06.051}
		{\bibfield  {journal} {\bibinfo  {journal} {Current Biology}\ }\textbf
			{\bibinfo {volume} {15}},\ \bibinfo {pages} {484} (\bibinfo {year}
			{2005})}\BibitemShut {NoStop}%
		\bibitem [{\citenamefont {Harris}(1978)}]{Harris1978}%
		\BibitemOpen
		\bibfield  {author} {\bibinfo {author} {\bibfnamefont {F.}~\bibnamefont
				{Harris}},\ }\href {\doibase 10.1109/PROC.1978.10837} {\bibfield  {journal}
			{\bibinfo  {journal} {Proceedings of the IEEE}\ }\textbf {\bibinfo {volume}
				{66}},\ \bibinfo {pages} {51} (\bibinfo {year} {1978})}\BibitemShut {NoStop}%
		\bibitem [{\citenamefont {Tukey}(1968)}]{Tukey1968}%
		\BibitemOpen
		\bibfield  {author} {\bibinfo {author} {\bibfnamefont {J.~W.}\ \bibnamefont
				{Tukey}},\ }\href@noop {} {\bibfield  {journal} {\bibinfo  {journal}
				{Spectral analysis of time series}\ }\textbf {\bibinfo {volume} {25}},\
			\bibinfo {pages} {25} (\bibinfo {year} {1968})}\BibitemShut {NoStop}%
		\bibitem [{\citenamefont {Winter}\ and\ \citenamefont
			{Chambon}(1986)}]{Winter1986}%
		\BibitemOpen
		\bibfield  {author} {\bibinfo {author} {\bibfnamefont {H.~H.}\ \bibnamefont
				{Winter}}\ and\ \bibinfo {author} {\bibfnamefont {F.}~\bibnamefont
				{Chambon}},\ }\href {\doibase 10.1122/1.549853} {\bibfield  {journal}
			{\bibinfo  {journal} {Journal of Rheology}\ }\textbf {\bibinfo {volume}
				{30}},\ \bibinfo {pages} {367} (\bibinfo {year} {1986})}\BibitemShut
		{NoStop}%
		\bibitem [{\citenamefont {Chambon}\ and\ \citenamefont
			{Winter}(1987)}]{Chambon1987}%
		\BibitemOpen
		\bibfield  {author} {\bibinfo {author} {\bibfnamefont {F.}~\bibnamefont
				{Chambon}}\ and\ \bibinfo {author} {\bibfnamefont {H.~H.}\ \bibnamefont
				{Winter}},\ }\href {\doibase 10.1122/1.549955} {\bibfield  {journal}
			{\bibinfo  {journal} {Journal of Rheology}\ }\textbf {\bibinfo {volume}
				{31}},\ \bibinfo {pages} {683} (\bibinfo {year} {1987})}\BibitemShut
		{NoStop}%
		\bibitem [{\citenamefont {Schultheisz}\ and\ \citenamefont
			{McKenna}(2000)}]{Schultheisz2000}%
		\BibitemOpen
		\bibfield  {author} {\bibinfo {author} {\bibfnamefont {C.~R.}\ \bibnamefont
				{Schultheisz}}\ and\ \bibinfo {author} {\bibfnamefont {G.~B.}\ \bibnamefont
				{McKenna}},\ }\href
		{https://www.nist.gov/publications/standard-reference-materials-non-newtonian-fluids-rheological-measurements}
		{\bibfield  {journal} {\bibinfo  {journal} {Proceeding Society of Plastics
					Engineers, Annual Technical Conference}\ }\textbf {\bibinfo {volume} {58}},\
			\bibinfo {pages} {1} (\bibinfo {year} {2000})}\BibitemShut {NoStop}%
		\bibitem [{\citenamefont {Tschoegl}(1989)}]{Tschoegl1989}%
		\BibitemOpen
		\bibfield  {author} {\bibinfo {author} {\bibfnamefont {N.~W.}\ \bibnamefont
				{Tschoegl}},\ }\href@noop {} {\emph {\bibinfo {title} {{The phenomenological
						theory of linear viscoelastic behavior}}}}\ (\bibinfo  {publisher}
		{Springer-Verlag Berlin Heidelberg},\ \bibinfo {year} {1989})\ p.\ \bibinfo
		{pages} {790}\BibitemShut {NoStop}%
		\bibitem [{\citenamefont {Koeller}(1984)}]{Koeller1984}%
		\BibitemOpen
		\bibfield  {author} {\bibinfo {author} {\bibfnamefont {R.~C.}\ \bibnamefont
				{Koeller}},\ }\href {\doibase 10.1115/1.3167616} {\bibfield  {journal}
			{\bibinfo  {journal} {Journal of Applied Mechanics}\ }\textbf {\bibinfo
				{volume} {51}},\ \bibinfo {pages} {299} (\bibinfo {year} {1984})}\BibitemShut
		{NoStop}%
		\bibitem [{\citenamefont {Scott-Blair}(1944)}]{Scott-Blair1944}%
		\BibitemOpen
		\bibfield  {author} {\bibinfo {author} {\bibfnamefont {G.~W.}\ \bibnamefont
				{Scott-Blair}},\ }\href {\doibase https://doi.org/10.1088/0950-7671/21/5/302}
		{\bibfield  {journal} {\bibinfo  {journal} {Journal of Scientific
					Instruments}\ }\textbf {\bibinfo {volume} {21}},\ \bibinfo {pages} {80}
			(\bibinfo {year} {1944})}\BibitemShut {NoStop}%
		\bibitem [{\citenamefont {Jaishankar}\ and\ \citenamefont
			{McKinley}(2012)}]{Jaishankar2012}%
		\BibitemOpen
		\bibfield  {author} {\bibinfo {author} {\bibfnamefont {A.}~\bibnamefont
				{Jaishankar}}\ and\ \bibinfo {author} {\bibfnamefont {G.~H.}\ \bibnamefont
				{McKinley}},\ }\href {\doibase 10.1098/rspa.2012.0284} {\bibfield  {journal}
			{\bibinfo  {journal} {Proceedings of the Royal Society A}\ }\textbf {\bibinfo
				{volume} {469}},\ \bibinfo {pages} {20120284, 1} (\bibinfo {year}
			{2012})}\BibitemShut {NoStop}%
		\bibitem [{\citenamefont {Caputo}(1995)}]{Caputo1995}%
		\BibitemOpen
		\bibfield  {author} {\bibinfo {author} {\bibfnamefont {M.}~\bibnamefont
				{Caputo}},\ }\href {\doibase 10.1007/BF02826009} {\bibfield  {journal}
			{\bibinfo  {journal} {Annali dell'Universit{\`{a}} di Ferrara}\ }\textbf
			{\bibinfo {volume} {41}},\ \bibinfo {pages} {73} (\bibinfo {year}
			{1995})}\BibitemShut {NoStop}%
		\bibitem [{\citenamefont {Jaishankar}\ and\ \citenamefont
			{McKinley}(2014)}]{Jaishankar2014}%
		\BibitemOpen
		\bibfield  {author} {\bibinfo {author} {\bibfnamefont {A.}~\bibnamefont
				{Jaishankar}}\ and\ \bibinfo {author} {\bibfnamefont {G.~H.}\ \bibnamefont
				{McKinley}},\ }\href {\doibase 10.1122/1.4892114} {\bibfield  {journal}
			{\bibinfo  {journal} {Journal of Rheology}\ }\textbf {\bibinfo {volume}
				{58}},\ \bibinfo {pages} {1751} (\bibinfo {year} {2014})}\BibitemShut
		{NoStop}%
		\bibitem [{\citenamefont {Winter}\ and\ \citenamefont
			{Mours}(1997)}]{Winter1997}%
		\BibitemOpen
		\bibfield  {author} {\bibinfo {author} {\bibfnamefont {H.~H.}\ \bibnamefont
				{Winter}}\ and\ \bibinfo {author} {\bibfnamefont {M.}~\bibnamefont {Mours}},\
		}in\ \href {\doibase 10.1007/3-540-68449-2_3} {\emph {\bibinfo {booktitle}
				{Advances in Polymer Science Neutron Spin Echo Spectroscopy Viscoelasticity
					Rheology}}},\ Vol.\ \bibinfo {volume} {134}\ (\bibinfo {year} {1997})\ pp.\
		\bibinfo {pages} {165--234}\BibitemShut {NoStop}%
		\bibitem [{\citenamefont {Braga}\ \emph {et~al.}(2006)\citenamefont {Braga},
			\citenamefont {Menossi},\ and\ \citenamefont {Cunha}}]{Braga2006}%
		\BibitemOpen
		\bibfield  {author} {\bibinfo {author} {\bibfnamefont {A.~L.}\ \bibnamefont
				{Braga}}, \bibinfo {author} {\bibfnamefont {M.}~\bibnamefont {Menossi}}, \
			and\ \bibinfo {author} {\bibfnamefont {R.~L.}\ \bibnamefont {Cunha}},\ }\href
		{\doibase 10.1016/j.idairyj.2005.06.001} {\bibfield  {journal} {\bibinfo
				{journal} {International Dairy Journal}\ }\textbf {\bibinfo {volume} {16}},\
			\bibinfo {pages} {389} (\bibinfo {year} {2006})}\BibitemShut {NoStop}%
		\bibitem [{\citenamefont {Gallot}\ \emph {et~al.}(2013)\citenamefont {Gallot},
			\citenamefont {Perge}, \citenamefont {Grenard}, \citenamefont {Fardin},
			\citenamefont {Taberlet},\ and\ \citenamefont {Manneville}}]{Gallot2013}%
		\BibitemOpen
		\bibfield  {author} {\bibinfo {author} {\bibfnamefont {T.}~\bibnamefont
				{Gallot}}, \bibinfo {author} {\bibfnamefont {C.}~\bibnamefont {Perge}},
			\bibinfo {author} {\bibfnamefont {V.}~\bibnamefont {Grenard}}, \bibinfo
			{author} {\bibfnamefont {M.-A.}\ \bibnamefont {Fardin}}, \bibinfo {author}
			{\bibfnamefont {N.}~\bibnamefont {Taberlet}}, \ and\ \bibinfo {author}
			{\bibfnamefont {S.}~\bibnamefont {Manneville}},\ }\href@noop {} {\bibfield
			{journal} {\bibinfo  {journal} {Review of Scientific Instruments}\ }\textbf
			{\bibinfo {volume} {84}},\ \bibinfo {pages} {045107, 1} (\bibinfo {year}
			{2013})}\BibitemShut {NoStop}%
		\bibitem [{\citenamefont {Saint-Michel}\ \emph {et~al.}(2016)\citenamefont
			{Saint-Michel}, \citenamefont {Gibaud}, \citenamefont {Leocmach},\ and\
			\citenamefont {Manneville}}]{SaintMichel2016}%
		\BibitemOpen
		\bibfield  {author} {\bibinfo {author} {\bibfnamefont {B.}~\bibnamefont
				{Saint-Michel}}, \bibinfo {author} {\bibfnamefont {T.}~\bibnamefont
				{Gibaud}}, \bibinfo {author} {\bibfnamefont {M.}~\bibnamefont {Leocmach}}, \
			and\ \bibinfo {author} {\bibfnamefont {S.}~\bibnamefont {Manneville}},\
		}\href {\doibase 10.1103/PhysRevApplied.5.034014} {\bibfield  {journal}
			{\bibinfo  {journal} {Physical Review Applied}\ }\textbf {\bibinfo {volume}
				{5}},\ \bibinfo {pages} {034014, 1} (\bibinfo {year} {2016})}\BibitemShut
		{NoStop}%
		\bibitem [{\citenamefont {Cipelletti}\ and\ \citenamefont
			{Weitz}(1999)}]{Cipelletti1999}%
		\BibitemOpen
		\bibfield  {author} {\bibinfo {author} {\bibfnamefont {L.}~\bibnamefont
				{Cipelletti}}\ and\ \bibinfo {author} {\bibfnamefont {D.~A.}\ \bibnamefont
				{Weitz}},\ }\href {\doibase 10.1063/1.1149894} {\bibfield  {journal}
			{\bibinfo  {journal} {Review of Scientific Instruments}\ }\textbf {\bibinfo
				{volume} {70}},\ \bibinfo {pages} {3214} (\bibinfo {year}
			{1999})}\BibitemShut {NoStop}%
		\bibitem [{\citenamefont {Blackman}\ and\ \citenamefont
			{Tukey}(1959)}]{Blackman1959book}%
		\BibitemOpen
		\bibfield  {author} {\bibinfo {author} {\bibfnamefont {R.~B.}\ \bibnamefont
				{Blackman}}\ and\ \bibinfo {author} {\bibfnamefont {J.~W.}\ \bibnamefont
				{Tukey}},\ }\href@noop {} {\emph {\bibinfo {title} {{The Measurement of Power
						Spectra from the Point of View of Communications Engineering}}}}\ (\bibinfo
		{publisher} {New York Dover},\ \bibinfo {year} {1959})\BibitemShut {NoStop}%
		\bibitem [{\citenamefont {Cheng}(2007)}]{Cheng2007}%
		\BibitemOpen
		\bibfield  {author} {\bibinfo {author} {\bibfnamefont {H.}~\bibnamefont
				{Cheng}},\ }\href@noop {} {\emph {\bibinfo {title} {{Advanced Analytical
						Methods in Applied Mathematics, Science and Engineering}}}}\ (\bibinfo
		{publisher} {LuBan Press},\ \bibinfo {year} {2007})\ p.\ \bibinfo {pages}
		{502}\BibitemShut {NoStop}%
		\bibitem [{\citenamefont {Bender}\ and\ \citenamefont
			{Orszag}(2013)}]{Bender2013}%
		\BibitemOpen
		\bibfield  {author} {\bibinfo {author} {\bibfnamefont {C.~M.}\ \bibnamefont
				{Bender}}\ and\ \bibinfo {author} {\bibfnamefont {S.~A.}\ \bibnamefont
				{Orszag}},\ }\href@noop {} {\emph {\bibinfo {title} {{Advanced Mathematical
						Methods for Scientists and Engineers I: Asymptotic Methods and Perturbation
						Theory}}}}\ (\bibinfo  {publisher} {Springer Science {\&} Business Media},\
		\bibinfo {year} {2013})\ p.\ \bibinfo {pages} {593}\BibitemShut {NoStop}%
	\end{thebibliography}
	%
	

\end{document}